\documentclass[twocolumn]{aastex62}
\pdfoutput=1 

\usepackage{amsmath,amstext}
\usepackage{listings}
\usepackage{threeparttable}
\usepackage[T1]{fontenc}
\usepackage{apjfonts} 
\usepackage[figure,figure*]{hypcap}
\usepackage{graphicx}
\usepackage{multirow}
\usepackage{xspace}
\usepackage{ulem}

\newcommand{\lxi}[1]{{\bf \color{blue} [Long: #1]}}

\newcommand{\tjg}[1]{{\bf \color{red} [TJG: #1]}}

\definecolor{royalpurple}{cmyk}{0.50, 0.60, 0.0, 0.3}

\newcommand*{\rom}[1]{\expandafter\@slowromancap\romannumeral #1@}
\newcommand{\rnum}[1]{\uppercase\expandafter{\romannumeral #1\relax}}

\newcommand{\chandra}{\textit{Chandra}\xspace}
\newcommand{\ciao}{\textit{CIAO}\xspace}

\newcommand{\kms}{\ensuremath{\;\mathrm{km\,s^{-1}}\xspace}
\newcommand{\km}{\ensuremath{\;\mathrm{km}}\xspace}
\newcommand{\pc}{\ensuremath{\;\mathrm{pc}}\xspace}
\newcommand{\kpc}{\ensuremath{\;\mathrm{kpc}}\xspace}
\newcommand{\AU}{\mbox{AU}\xspace}
\newcommand{\s}{\ensuremath{\mathrm{s}}\xspace}
\newcommand{\yr}{\ensuremath{\mathrm{yr}}\xspace}
\newcommand{\Myr}{\ensuremath{\mathrm{Myr}}}\xspace}
\newcommand{\degree}{$^{\circ}$\xspace}

\newcommand{\msol}{\ensuremath{\mbox{$\mathrm{M}_{\sun}$}}\xspace}

\shorttitle{AASTeX 6.2 Template}
\shortauthors{Xi et al.}

\begin{document}

\title{The Expansion of the Forward Shock of 1E 0102.2-7219 in X-rays}

\author{Long Xi \href{https://orcid.org/0000-0003-3350-1832}{\includegraphics[scale=0.01]{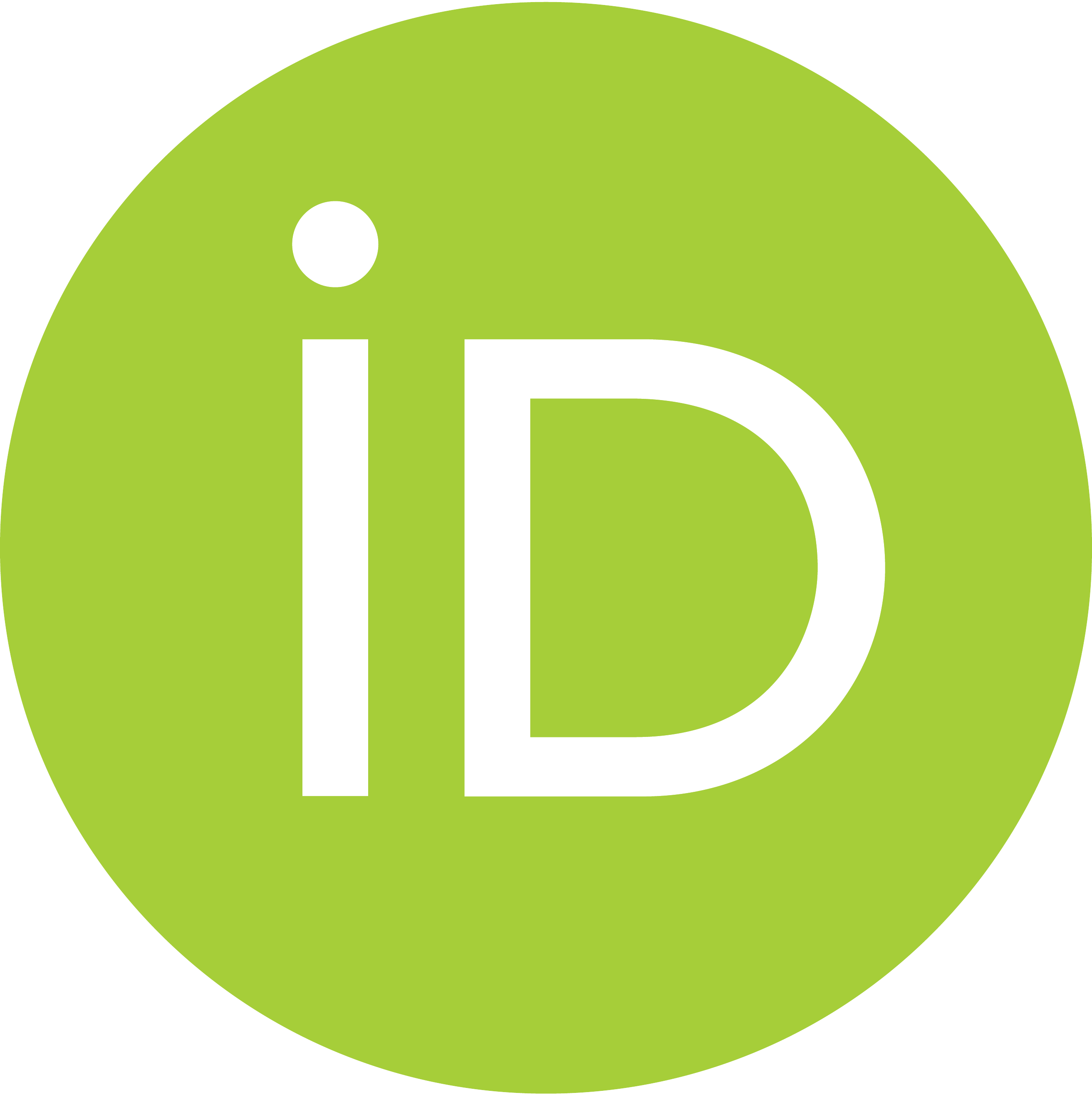}}}
\affiliation{Key Laboratory for Particle Astrophysics, Institute of High Energy Physics, Chinese Academy of Sciences,Beijing 100049, P.R. China;longx@ihep.ac.cn}
\affiliation{University of Chinese Academy of Sciences,Beijing 100049, P.R. China;}
\author{Terrance J. Gaetz \href{https://orcid.org/0000-0002-5115-1533}{\includegraphics[scale=0.01]{orcid.pdf}}}
\affiliation{Harvard-Smithsonian Center for Astrophysics, 60 Garden St., Cambridge, MA 02138, USA; tgaetz@cfa.harvard.edu, pplucinsky@cfa.harvard.edu, dpatnaude@cfa.harvard.edu}
\author{Paul P. Plucinsky \href{https://orcid.org/0000-0003-1415-5823}{\includegraphics[scale=0.01]{orcid.pdf}}}
\affiliation{Harvard-Smithsonian Center for Astrophysics, 60 Garden St., Cambridge, MA 02138, USA; tgaetz@cfa.harvard.edu, pplucinsky@cfa.harvard.edu, dpatnaude@cfa.harvard.edu}
\author{John P. Hughes
\href{https://orcid.org/0000-0002-8816-6800}{\includegraphics[scale=0.01]{orcid.pdf}}}
\affiliation{Department of Physics and Astronomy, Rutgers University, 136 Frelinghuysen Road, Piscataway, NJ 08854-8109, USA; jph@physics.rutgers.edu}
\author{Daniel J. Patnaude
\href{https://orcid.org/0000-0002-7507-8115}{\includegraphics[scale=0.01]{orcid.pdf}}}
\affiliation{Harvard-Smithsonian Center for Astrophysics, 60 Garden St., Cambridge, MA 02138, USA; tgaetz@cfa.harvard.edu, pplucinsky@cfa.harvard.edu, dpatnaude@cfa.harvard.edu}

\begin{abstract}

We measure the expansion of the forward shock of the Small Magellanic Cloud supernova remnant 1E\,0102.2-7219 in X-rays using {\em Chandra X-Ray Observatory} on-axis \textit{Advanced CCD Imaging Spectrometer} (ACIS) observations from 1999--2016. We estimate an expansion rate of $0.025\%\pm0.006\%~\mathrm{yr}^{-1}$ and a blast-wave velocity of $1.61\pm0.37\times10^3~\mathrm{km~s}^{-1}$. Assuming partial electron-ion equilibration via Coulomb collisions and cooling due to adiabatic expansion, this velocity implies a postshock electron temperature of $0.84\pm0.20~\mathrm{keV}$ which is consistent with the estimate of $0.68\pm0.05~\mathrm{keV}$ based on the X-ray spectral analysis. We combine the expansion rate with the blast wave and reverse shock radii to generate a grid of one-dimensional models for a range of ejecta masses ($2-6\,\msol$) to constrain the explosion energy, age, circumstellar density, swept-up mass, and unshocked-ejecta mass. We find acceptable solutions for a constant density ambient medium and for an $r^{-2}$ power-law profile (appropriate for a constant progenitor stellar wind). For the constant density case, we find an age of $\sim\,1700~\mathrm{yr}$, explosion energies $0.87-2.61\times10^{51}~\mathrm{erg}$, ambient densities $0.85-2.54~\mathrm{amu\,cm}^{-3}$, swept-up masses $22-66\,\msol$, and unshocked-ejecta masses $0.05-0.16\,\msol$. For the power-law density profile, we find an age of $\sim 2600~\mathrm{yr}$, explosion energies $0.34-1.02\times10^{51}~\mathrm{erg}$, densities $0.22-0.66~\mathrm{amu\,cm}^{-3}$ at the blast wave, swept-up masses $17-52\,\msol$, and unshocked-ejecta masses $0.06-0.18\,\msol$. Assuming the true explosion energy was $0.5-1.5\times10^{51}~\mathrm{erg}$, ejecta masses $2-3.5\,\msol$ are favored for the constant density case and $3-6\,\msol$ for the power-law case. The unshocked-ejecta mass estimates are comparable to Fe masses expected in core-collapse supernovae with progenitor mass $15.0-40.0\,\msol$, offering a possible explanation for the lack of Fe emission observed in X-rays.

\end{abstract}


 \section{Introduction}
 \label{sec:introduction}

\begin{figure*}[htbp]
  \centering
  \includegraphics[width=1.0\textwidth]{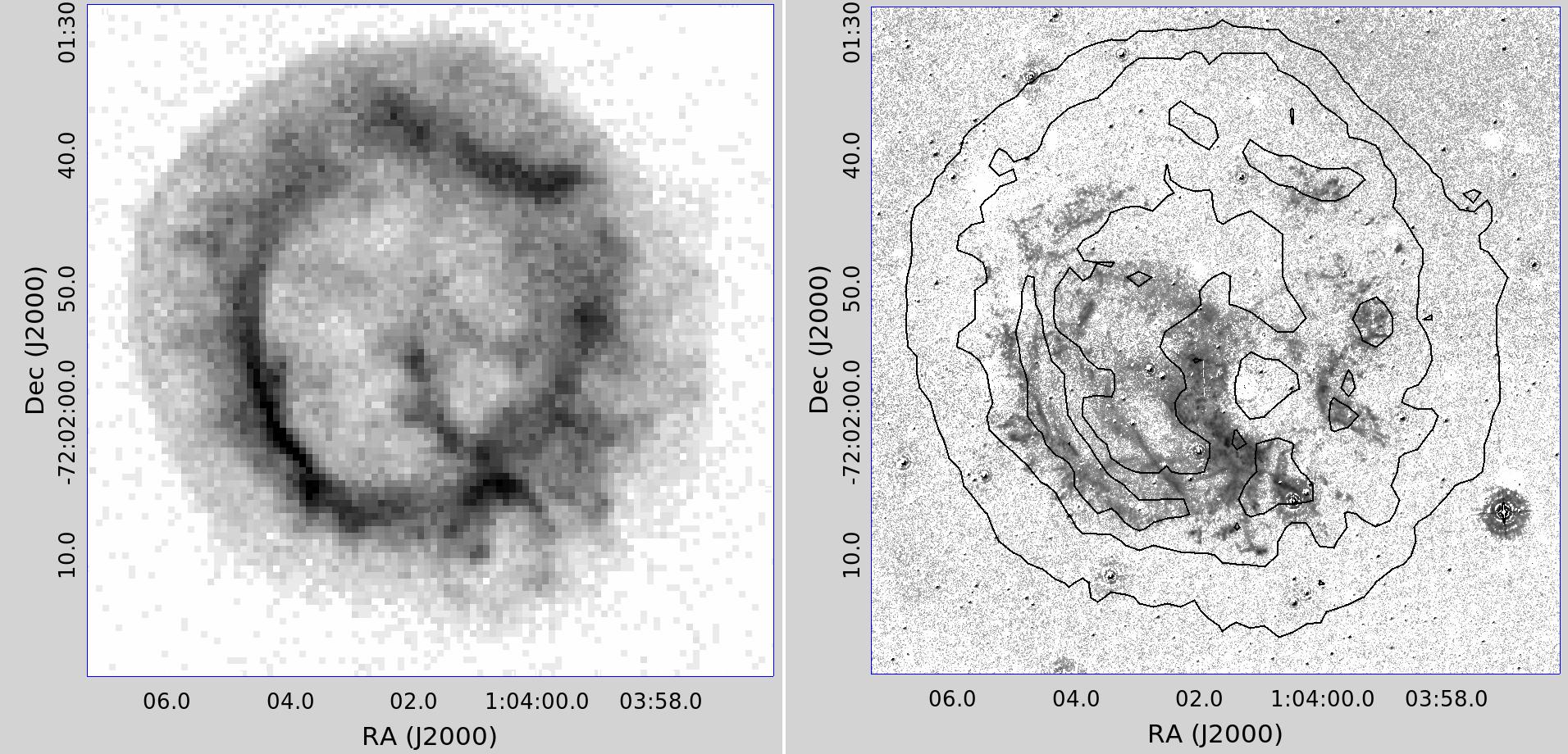} 
   \caption{X-ray and optical emission of E0102. Left panel: {\em Chandra} data for \texttt{ObsID} 1423 (binned to 1 ACIS sky pixel, 0.4-2.0\,keV), with a sqrt color stretch. Right panel: HST ACS 
   data (continuum subtracted) with a logarithmic color stretch. X-ray contours are overplotted on the optical image.}
   \label{fig:opt.xray.two.panel}
\end{figure*}

%
%
The supernova remnant (SNR) 1E~0102.2-7219 (hereafter E0102) is the X-ray brightest SNR in the  Small Magellanic Cloud (SMC) with a luminosity of $1.1\times10^{37}$ $\mathrm{erg}\,\mathrm{s}^{-1}$ in the 0.5-2.0~keV band.  E0102 was discovered by \citet{seward1981} with the {\em Einstein Observatory}.  It was classified as an ``O-rich'' SNR based on the
optical spectra acquired soon after the X-ray discovery by \citet{dopita1981} and confirmed by follow-up observations of
the complex optical emission in [\ion{O}{3}] and [\ion{O}{2}] \citep{tuohy1983}.
\cite{blair1989} presented the first UV spectra of E0102 and argued for a progenitor mass between 15 and 25\,\msol based on the derived O, Ne, and Mg abundances.  \cite{blair2000} refined this argument with {\em Wide Field and Planetary Camera 2} (WFPC2) and {\em Faint Object Spectrograph} data from {\em  Hubble Space Telescope} (HST) to suggest that the precursor was a Wolf-Rayet star of between  25 and 35\,\msol with a large O mantle that produced a Type Ib supernova.  The  optical morphology  shows a complicated, filamentary structure first seen by \cite{tuohy1983} and then seen in more detail
in the HST WFPC2 [\ion{O}{3}] $\lambda=5007$ (F502N) image in  \citet{blair2000}.
\citet{finkelstein2006} confirm this  complicated structure with  images from several filters from the  {\em Advanced Camera for Surveys} (ACS) on  HST.  In contrast, the X-ray morphology  observed  with the  {\em Chandra X-ray Observatory} ({\em Chandra})  is considerably simpler than  that observed in the optical.  \citet{gaetz2000} presented  high resolution  images from  {\em Chandra} that allow the  blast wave emission  to be separated from the  ejecta emission (see Figure~\ref{fig:opt.xray.two.panel}). There is a good correlation between the optical and X-ray emission for some of the filaments as shown in the right panel of Figure~\ref{fig:opt.xray.two.panel} but for other filaments there is little or no correlation.  The outer blastwave is well-defined in the X-rays but is notably absent in the optical. 

In an early {\em Chandra} study of the first ACIS images of
E0102 (\texttt{ObsID} 1231), \citet[][hereafter HRD00]{hughes2000}
investigated the post-shock partition of energy among electrons, ions,
and a putative population of relativitisic particles, i.e., cosmic
rays, at the forward shock of the remnant.  The forward shock was
cleanly resolved in the ACIS data and its spectrum was extracted;
using nonequilibrium ionization (NEI) thermal emission models, HRD00
determined the post-shock electron temperature to be 0.4-1 keV with a
significant component of the acceptable range due to model
uncertainty.  The proper motion of the entire SNR (assumed to be both
radially and azimuthally uniform) was determined by comparing the
{\em Chandra} ACIS image to archival {\em Einstein} and 
{\em R\"ontgensatellite} ({\em ROSAT}) High Resolution
Imager images taken $\sim$8 and $\sim$20 years earlier, respectively.
These earlier images had 50\% encircled energy radii of
$4^{\prime\prime}$, more than 10 times larger than that of the
{\em Chandra} ACIS image.  The fractional expansion rate of the entire
remnant was $0.100\% \pm 0.025\%$ yr$^{-1}$, which, when extrapolated
to the location of the forward shock, implied a shock speed of
$\sim$6000 km s$^{-1}$. Converting the shock speed to an electron
temperature required accounting for the uncertain amount of anomalous
heating at the shock front followed by energy exchange through Coulomb
collisions between the electron and ion thermal populations, as well
as estimating the amount of adiabatic decompression in the measured
postshock region.  After these calculations, HRD00 found the minimum
expected electron temperature to be $\sim$2.5 keV, significantly
higher than the measured temperature value, leading to the suggestion
that some of the forward shock energy in E0102 went into generating
cosmic rays.


In this paper we make the first direct measurement of
the forward shock speed in E0102 using multiple epochs of
{\em Chandra} observations.  We use {\em Chandra}'s exquisite angular
resolution to separate the forward shock from the interior emission
of the remnant. This allows us to measure the expansion of the forward
shock alone without any confusion from other parts of the remnant that
may have different velocities. We closely follow the approach of HRD00 while updating
the electron temperature measurements, improving the shock electron
temperature model, and constraining the age, ambient medium density,
ejected mass, and other dynamical quantities from an analytical shock
model.  Our paper is structured as follows. In
\S \ref{sec:chandra.observations.e0102} we describe the {\em Chandra} ACIS 
observations of E0102 and
in \S\ref{sec:analysis} we present our full analysis including measuring the proper
motion of the forward shock and the radial locations of the forward
and reverse shocks. \S\ref{sec:one.dimensional.shock.models}  
describes our one-dimensional analytical shock model and gives results on the
dynamical state of the SNR. Also in Section \S\ref{sec:one.dimensional.shock.models}  we determine the post-shock
electron temperature directly from spectral modeling and also by
calculation from the forward shock speed. \S\ref{sec:conclusion}  concludes.  Throughout this
paper we assume the distance to the SMC is 60.6 kpc \citep{hilditch2005}.
Error bars in plots and uncertainties on numerical values 
are quoted at the 1$\sigma$ confidence level unless otherwise stated.

\clearpage
\section{{\em Chandra} Observations of E0102}
\label{sec:chandra.observations.e0102}
As a calibration source for {\em Chandra}, E0102 has been observed every year from 1999 to 2018. An overview of {\em Chandra} including the imaging capabilities of the {\em High Resolution Mirror Assembly}~(HRMA) is given in \citep{weiss2000,weiss2002}.  We utilize data from the {\em Advanced CCD Imaging Spectrometer}~(ACIS) instrument which is described in detail in ~\cite{garmire92,bautz98,garmire03}. The ACIS instrument contains two arrays of charge-coupled devices (CCDs), an imaging array called ACIS-I and a spectroscopy array called ACIS-S.  The central CCD of the ACIS-S array (called S3)  is a backside-illuminated (BI) CCD which has superior low energy quantum efficiency compared to the frontside-illuminated (FI) CCDs  in ACIS.
We select 12 ACIS-S3 observations, with exposure times ranging from $\sim$ 7.6-19.7~ks (see Table~\ref{tab:obslist}), that had the center of the remnant positioned within $1^{\prime}$ of the on-axis aim point to provide the optimal imaging performance with {\em Chandra}. We exclude from our analysis the off-axis observations of E0102 with sub-optimal imaging and the on-axis observations after 2017 as the correction for the additional absorption caused by the contamination layer on the ACIS filters has become less certain after 2017 \citep{plucinsky2018}. Each observation is reprocessed to generate a new level=2 event list, using \ciao version 4.8 and CALDB 4.7.2, by the \ciao tool \texttt{chandra\_repro}, with the option \texttt{pix\_adj} set to the default, such that the {\em Energy Dependent Sub-pixel Event Reconstruction} (\texttt{EDSER}) algorithm is applied. This enables sub-pixel resolution in imaging data analysis \citep{li2004} by using the distribution of charge amongst the pixels within a $3\times 3$ pixel detection island to estimate a better position for where the X-ray landed in the central pixel.

E0102 was observed in ACIS subarray mode starting in 2003 to achieve a shorter frametime which reduces photon ``pileup''.  Photon pileup is the condition in which two or more photons interact with the CCD within the event detection cell (typically $3\times3$ pixels for ACIS) within a single frame, resulting in an incorrect energy and perhaps position for the events.  The brightest parts of E0102 (for example the white and red regions in Figure~\ref{fig:regions.pileup.map} bottom left panel) are bright enough to produce significant pileup which depresses the observed count rate and shifts the distribution of observed energies to higher energies.  The pileup level can exceed 5\% as shown for a fullframe observation with a 3.2\,s frametime (bottom left panel of Figure~\ref{fig:regions.pileup.map}). Pileup can be significantly reduced if the frametime is decreased to 0.8s as it was in the 2003 and later observations.  The outer blastwave has a low enough surface brightness that pileup is negligible even in the earliest observations in fullframe mode. Another factor which reduces pileup in the later E0102 observations is the accumulation of a contamination layer on the ACIS optical blocking filters (OBF) \citep[see][]{plucinsky2003,plucinsky2016}.  The ACIS contamination layer produces a highly energy-dependent absorption with low energy photons around 0.5~keV suffering high absorption while higher energy photons around 1.5~keV and up suffering little absorption. By the time of the observation in 2016, the combination of the shorter frametime and the contamination layer resulted in the bright parts of the ring having negligible pileup.
 
 For our analysis, we focus primarily on the outer blastwave which is free from the effects of pileup even for the fullfame mode observations. We make use of only two fullframe mode observations, one from 1999 (\texttt{ObsID} 1423) as the first observation in a sequence to measure the position of the blastwave and the second from 2006 (\texttt{ObsID} 6766).  We include  6766 as a check on our registration method discussed in \S\ref{sec:img_reg} as it is the latest fullframe observation on S3. The remaining observations in our analysis are in subarray mode (see Table~\ref{tab:obslist}) and in fact all of the observations after 2006 are in subarray mode.

\begin{table}[htb]
\caption{Observation List}
\begin{center}
\label{tab:obslist}
\begin{tabular}{ l l c c c}
\hline\hline
ObsID & Start Date & Exposure & Time difference &Full-frame\\ & & (ks) & (yr) &\\
\hline 
\phantom{0}1423 & 1999 Nov 01 & 18.92 & \phantom{0}0.00 & Y\\
\phantom{0}3545 & 2003 Jun 06 & \phantom{0}7.86 & \phantom{0}3.60 & N\\
\phantom{0}6765 & 2006 Mar 19 & \phantom{0}7.64 & \phantom{0}6.38 & N\\
\phantom{0}6766 & 2006 Jun 06 & 19.70 & \phantom{0}6.59 & Y\\
\phantom{0}9694 & 2008 Feb 07 & 19.20 & \phantom{0}8.27& N\\
11957 & 2009 Dec 30 & 18.45 & 10.17 & N\\
13093 & 2011 Feb 01 & 19.05 & 11.26 & N\\
14258 & 2012 Jan 12 & 19.05 & 12.20 & N\\
15467 & 2013 Jan 28 & 19.08 & 13.25& N\\
16589 & 2014 Mar 27 & \phantom{0}9.57 & 14.41& N\\
17380 & 2015 Feb 28 & 17.65 & 15.34 & N\\
18418 & 2016 Mar 15 & 14.33 & 16.38 & N\\
\hline
\end{tabular}
\end{center}
\end{table}

\section{Analysis}
\label{sec:analysis}

The objective of our analysis is to measure the change in the position of the blast wave of the SNR as a function of time to determine the velocity of the shock. Our approach is summarized here and described in detail in the following sections.  We first construct a reference image (called the ``model'') from an observation early in the {\em Chandra} mission that had a relatively high count rate to provide good statistics. We construct images from the later observations (called the ``comparison data'') using the same processing as the reference image.  We correct the model image for the change in quantum efficiency (QE) as a function of time and energy given the time difference between the observation date of the model image and that of the comparison data. We register the images using the bright, central feature in E0102.  We extract radial profiles in 16 directions and fit the profiles with a model of the shock emission smoothed by the angular resolution of the combined HRMA and ACIS system. We measure the shifts between the model and comparison data profiles to derive offsets in the 16 directions.  We then fit the offsets as a function of position angle with a sinusoidal function to account for any remaining error in the registration.  From this fit, we determine the best fitted value of the average expansion for each combination of comparison data and model image. We then repeat this exercise for the subarray observations in Table~\ref{tab:obslist} 
using \texttt{ObsID} 1423 for the reference image
to determine the average expansion as a function of time.

\subsection{Image Generation}
\label{subsec:image.generation}
The images were created by using the standard \ciao tool \texttt{dmcopy} to bin the events into $0.246\arcsec\times0.246\arcsec$ sky pixels in the $0.5-2.0$~keV energy range.
The \texttt{EDSER} algorithm had already been applied to the events list before the images were created. The imaging improvement results from dither moving the pixel center across the image combined with the slight imaging improvement from \texttt{EDSER}.
The pixel size of $0.246\arcsec\times0.246\arcsec$ was selected to optimize the imaging information of the HRMA and ACIS combination while providing sufficient statistics in a single pixel. The energy range of  $0.5-2.0$~keV was selected to optimize the signal from the SNR compared to the detector and sky background components. 

\subsubsection{Model Image Generation}
\label{subsubsec:model.image.generation}
\texttt{ObsID} 1423 was used to construct the model image of E0102 since the exposure time of 18.9~ks was relatively long compared to other observations of E0102 and the observation was executed on 1999 November 01 when the ACIS contamination layer was relatively thin. Because the quantum efficiency (QE) and exposure time changed between \texttt{ObsID} 1423 and later observations, the counts in each bin of the model image need to be weighted by a factor to account for the QE and exposure time differences.   The weighting factor was applied to each event in the events list of \texttt{ObsID} 1423 and a new image of the weighted events was created using \texttt{dmcopy}. The weights were calculated as $q_2 t_2 / q_1 t_1$ where $q_1$ is the QE at the energy of the event for the date of \texttt{ObsID} 1423, $q_2$ is the QE for the date of the comparison observation, $t_1$ is the exposure time of \texttt{ObsID} 1423, and $t_2$ is the exposure time of the comparison observation. To determine the QE of an event at a given energy and a given point in time, we used the \ciao tool \texttt{eff2evt} by setting the option \texttt{detsubsysmod} to the start time of the observation. We obtained $q_{1}$ by setting \texttt{detsubsysmod} to the start time of \texttt{ObsID} 1423, and $q_{2}$ by setting it to the start time of comparison observation. The model image was smoothed with a $\sigma=0.492\arcsec$ Gaussian.
Figure~\ref{fig:regions.pileup.map} displays the image from \texttt{ObsID} 1423 in the $0.5-2.0$\,keV band in the top panel.  The lower left panel shows the pileup percentage as calculated by the \ciao tool \texttt{pileup\_map} for \texttt{ObsID} 1423, and the bottom right panel shows the pileup percentage for a subarray mode
observation \texttt{ObsID} 3545 (note that the color scales for the bottom left and bottom right panels are different).  Since \texttt{ObsID} 1423 used a frametime of 3.2\,s and the contamination layer was relatively thin in 1999, the bright ring of E0102 has significant pileup (maximum value is $\sim10\%$).  However, the outer blastwave is essentially free of pileup given that it is much fainter than the bright ring. 


\iftrue
\begin{figure*}[htbp]
\centering
\begin{tabular}{ll}
\includegraphics[width=0.49\textwidth]{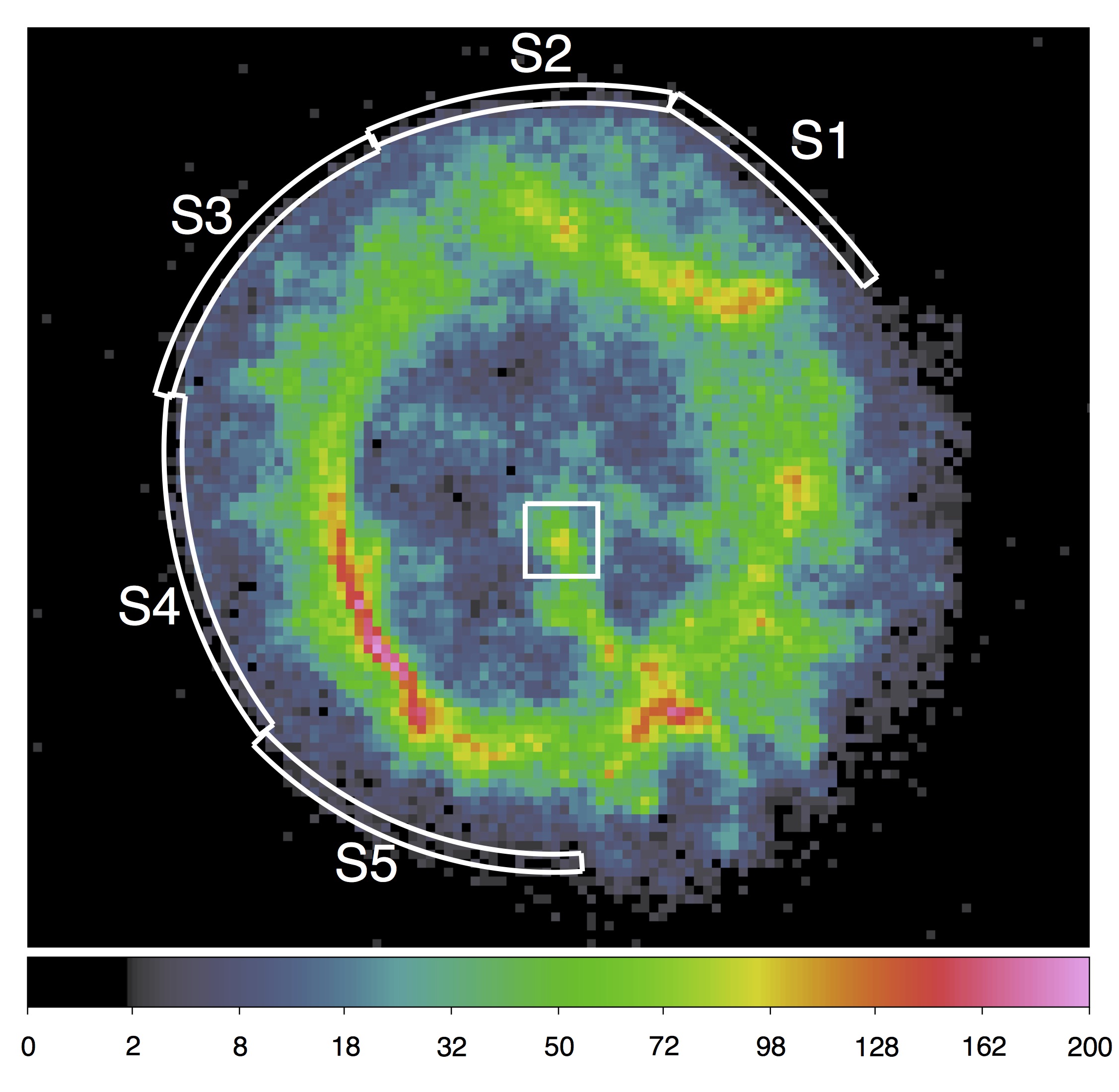} & \\
\includegraphics[width=0.49\textwidth]{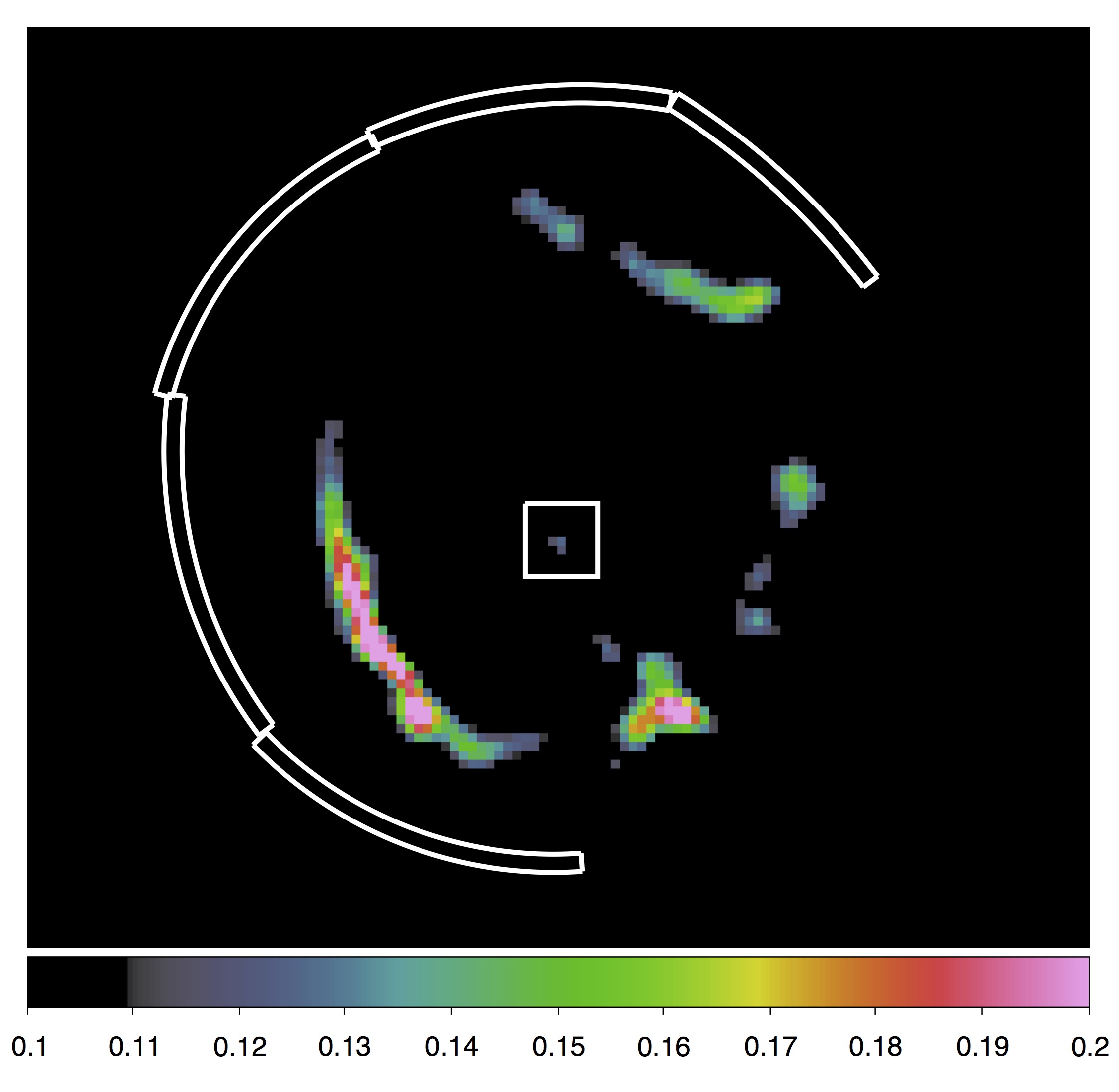} &
\includegraphics[width=0.49\textwidth]{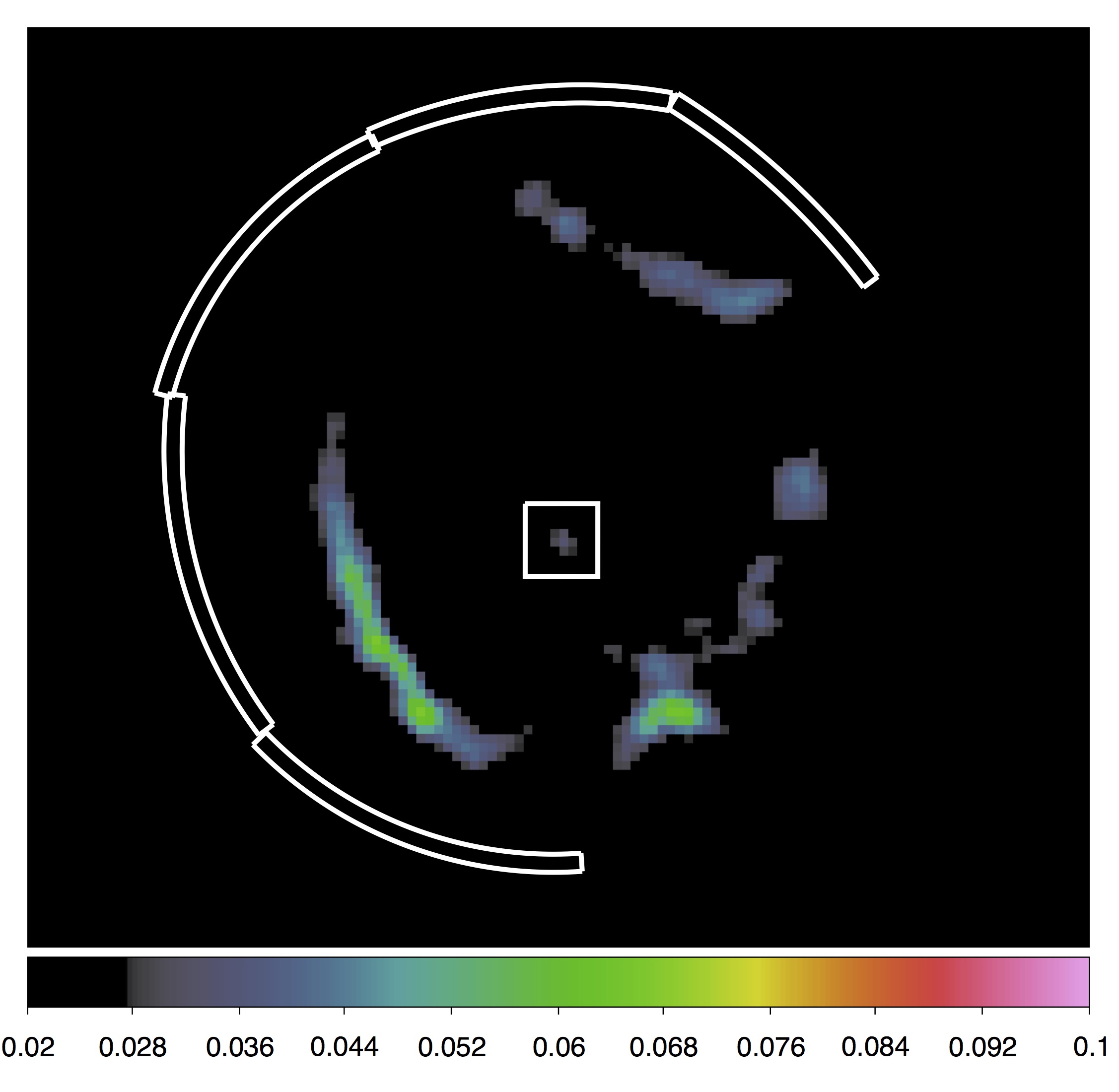}
\end{tabular}
\caption{Top Left: Image from \texttt{ObsID}~1423 in the 0.5--2.0~keV band. The units for the image and the colorbar are counts. The annular regions ($1.0\arcsec$ width) show the spectral extraction regions for the blastwave spectrum and the box region in the center shows the region used for registration. 
Bottom Left: \texttt{ObsID} 1423 pileup map (early full-frame observation). The units for the image and color bar are counts/frame.  A value of 0.1/0.2 counts/frame corresponds to $\sim5/10\%$ pileup. Bottom Right: pileup map of \texttt{ObsID} 3545 in 2003, a subarray observation with a shorter frametime.  The units for the image and color bar are counts/frame. Pileup is less than $5\%$ even for the brightest regions in \texttt{ObsID} 3545. Note that the color scale ranges for the bottom panels are different.
}
\label{fig:regions.pileup.map}
\end{figure*}
\fi

\subsection{Comparison Data Image Generation}
\label{subsec:comparison.data.image.generation}
The subarray observations in Table~\ref{tab:obslist} were used to construct the comparison data images. 
Figure~\ref{fig:regions.pileup.map} (bottom right panel) shows the image from \texttt{ObsID}~3545 with a 1.1\,s frametime from 2003 when the contamination layer was thicker than it was for \texttt{ObsID}~1423.  Since the frametime is shorter than it was for \texttt{ObsID}~1423 and the rate incident on the detector is reduced due to the contamination layer,
the pileup is significantly reduced with no pixel having more than 5\% pileup. Note that the color scale is different for this image than for the pileup map for \texttt{ObsID}~1423.
All subarray observations listed in Table~\ref{tab:obslist} observed after ObsID 3545  use a frametime of 0.8~s, which reduces pileup further.
The comparison images have no corrections applied to them and are treated as the ``data''.  As discussed in the previous section, the image from \texttt{ObsID}~1423 is treated as the ``model'' and hence the QE and exposure time corrections are applied to that image before comparison with the later data sets.

\subsection{Image Registration}
\label{sec:img_reg}
Since the observations of E0102 after 2006 were acquired in subarray mode, there are no point sources in the field of view which we can use for registration. To take advantage of the nearly 17 year baseline of observations of E0102, we used the central bright feature defined by a $4^{\prime\prime}$ by $4^{\prime\prime}$ box region (shown in Figure~\ref{fig:regions.pileup.map}, top panel), centered on  $\alpha_{J2000}=01^{h}04^{m}2.1^{s}$, $\delta_{J2000}=-72^{\circ}1^{\prime}55.6^{\prime\prime}$ to register the observations to \texttt{ObsID}~1423.  This feature, which is near the expansion center of the optical filaments given by \citet[][hereafter F06]{finkelstein2006}, is located close to some of the bright optical filaments shown in Figure~\ref{fig:opt.xray.two.panel}. According to measurements by \citet{eriksen2001} and \citet{Vogt2010}, this region contains blue-shifted material with a maximum velocity ranging from 2100$\;\mathrm{km}\,\mathrm{s}^{-1}$ to as high as 2500$\;\mathrm{km}\,\mathrm{s}^{-1}$.  Comparing these velocities to the average bulk motion velocity of 1966 $\mathrm{km}\,\mathrm{s}^{-1}$ of the filaments surrounding the center in F06 suggests that the central bright feature is moving along the line-of-sight direction; thus it has little or no proper-motion component. 

In addition, there is no evidence for the existence of a central compact object in this region; \citet{rutkowski2010} place an upper limit on the $0.1-10.0$~keV band flux of $\sim7.0\times10^{34}\mathrm{erg~s}^{-1}$ depending on the assumed spectrum of a possible point source.  This rules out the existence of a high-velocity neutron star coincident with this feature. \citet{vogt2018} claim the detection of a point source in X-rays at a different position,  $\alpha_{J2000}=01^{h}04^{m}2.7^{s}$, $\delta_{J2000}=-72^{\circ}2^{\prime}00.2^{\prime\prime}$, about $5.0^{\prime\prime}$ away from the central knot, which they conclude is a neutron star similar to the Compact Central Object (CCO) in Cas~A and other remnants.
Based on these results, we assume that the central bright feature in X-rays has no or unmeasurable proper-motion (although there may be a relatively large velocity component along the line-of-sight), and can therefore be used to register the images of E0102 from different epochs to each other.  

\subsubsection{Registration Method and Results}
\label{subsubsec:registration.method.and.results}

We adopt a registration method similar to that described in \citet{vink2008}. To obtain the shift between \texttt{ObsID} 1423 and later observations, we calculated the C statistic \citep{cash1979}, which is the maximum likelihood statistic for a Poisson distribution, for the $4\arcsec$ by $4\arcsec$ central bright feature in the model image and comparison data image by;
\begin{equation} \label{eq:1}
C = 2 \sum^{N}_{i,j=1} m_{i,j} - n_{i,j} + n_{i,j}(\ln(n_{i,j})-\ln(m_{i,j}));
\end{equation}
where, $n_\mathit{i,j}$ is the counts in bin $(i,j)$ of comparison data image, $m_\mathit{i,j}$ is the expected counts in bin $(i,j)$ of model image. The use of the C statistic is justified by the low number of counts per pixel. For example, for a late observation such as \texttt{ObsID}~17380 the distribution of counts per pixel from the central region has a shape similar to a Poisson distribution with a mode of $\sim3.7$ counts per pixel and minimum and maximum values of 0 and 14 counts per pixel.
We shift the events in the events list of \texttt{ObsID}~1423 by steps of $0.0246^{\prime\prime}$ ($\frac{1}{10}$ the pixel size in the image), regenerate the model image, and then recalculate the C statistic of the new model image and data image. In this manner, we generate a two dimensional distribution of C statistic values versus shifted positions of the model image. We fit this two dimensional distribution to determine the offset in $x$ and $y$ that minimizes the C statistic. We fit the C statistic values versus shift positions with a quartic function around the minimum of the C statistic ($C_\mathit{min}$) in both the $x$ and $y$ direction. For each direction, there are 13 data points included in the fit, the point at $C_\mathit{min}$ and 6 points on either side. The quartic function was used to account for the possibility of an asymmetric distribution of the C statistic with $x$ and/or $y$. We also tried a quadratic function but the fits were poor given the asymmetric distributions and a sextic function but the results were nearly identical to the quartic function.
We adopt the shift positions of the minimum values of the quartic curves in the $x$ and $y$ directions as the shift of \texttt{ObsID} 1423 with respect to the later observation. This shift is then used to register \texttt{ObsID} 1423 to the later observation.

Table~\ref{tab:2} lists the shifts that were determined for the registration of the comparison data sets to the model image. 
The mean shift in the X direction is $0.20\arcsec$ with an uncertainty of $0.06\arcsec$ and the mean shift in the Y direction is $0.18\arcsec$ with an uncertainty of $0.07\arcsec$.
Shifts on the order of $0.3\arcsec$ are consistent with the accuracy of the {\em Chandra} aspect reconstruction. \chandra dithers during observations, executing a Lissajous pattern within a $16\arcsec\times16\arcsec$ box for ACIS observations.  The aspect reconstruction must account for this dither pattern as a function of time when assigning coordinates to each event.  The absolute astrometry for \chandra observations can be improved by using the known positions of X-ray sources if the positions are known to high precision.  Unfortunately we are not able to apply this technique to our subarray data because of the lack of sources in the field-of-view.  The most accurate absolute astrometry is not necessary for our analysis, rather, we need the most accurate relative astrometry amongst our observations.

\begin{table}[htbp]
\caption{Registration Shift Results}
\begin{center}
\label{tab:2}
\begin{tabular}{ l c c c c}
\hline\hline
ObsID & shift x & shift y &  error x & error y\\
&[arcsec]&[arcsec]&[arcsec]&[arcsec]\\
\hline 
\phantom{0}3545 & -0.18 & -0.10 & 0.06 & 0.11\\
\phantom{0}6765 & -0.26 & -0.06 & 0.06 & 0.06\\
\phantom{0}9694 & -0.19 & -0.06 & 0.04 & 0.03\\
11957 & -0.19 & \phantom{-}0.19 & 0.04 & 0.04\\
13093 & -0.19 & -0.26 & 0.03 & 0.06\\
14258 & \phantom{-}0.06 & -0.02 & 0.04 & 0.04\\
15467 & -0.01 & \phantom{-}0.17 & 0.04 & 0.06\\
16589 & -0.21 & -0.20 & 0.08 & 0.14\\
17380 & -0.25 & \phantom{-}0.23 & 0.06 & 0.06\\
18418 & -0.20  & 0.48 & 0.07 & 0.10\\
%
%
\hline
\end{tabular}
\end{center}
\end{table}

\subsubsection{Registration Systematic Uncertainty}
\label{subsubsec:reg.sys.uncertainty}
The statistical error in the registration is estimated by the change in the C statistic, $\Delta C = C - C_\mathit{min}$, to define confidence intervals.  We adopt a $\Delta C = 1.0$ as the equivalent of a Gaussian $1\sigma$ uncertainty. To estimate the systematic uncertainty of our registration method, we applied it to measure the shift between \texttt{ObsID} 1423 and four full frame observations from 2003 to 2006 listed in Table~\ref{tbl:3}. 
The full frame data have the advantage that there are point sources bright enough in the field-of-view that can be used to register the images. We can register two observations by our method using the bright central feature and then register the same two observations using the point sources. We can then compare the results for consistency. We used the \ciao tool \texttt{wavdetect} to identify point sources with a significance threshold of $10^{-6}$ for spatial scales 2, 4, and 8. 
The input image for {\tt wavdetect} was a 0.35--7.0\,keV image (bin size 0.5 sky pixel) generated by the \ciao tool \texttt{fluximage}. 
A point-spread function (PSF) map was generated using the \ciao tool \texttt{mkpsfmap} for 0.92\,keV and an enclosed counts fraction (\texttt{ecf}) of 0.393. 
The locations of the point sources were refined using an iterative $\sigma$-clipping algorithm. Events were filtered to 0.35--7.0\,keV and an initial clipping radius of 5 ACIS sky pixels centered on the original position estimate was used.  The centroids of the events within that radius were evaluated and events with distance greater than 3$\sigma$ from the centroid were excluded. This process was performed for 10 iterations, or until the centroid changed by less than 0.01 sky pixel.

\begin{table}[htbp]
\caption{Observations Used in Estimating Registration Uncertainty}
\begin{center}
\label{tbl:3}
\begin{tabular}{ l l c c}
\hline\hline
ObsID & Start Date & Exposure time & Full-frame \\ & & [ks] &  \\
\hline 
1423 & 1999 Nov 1 & 18.92 & Y\\
5123 & 2003 Dec 15 & 20.32 & Y\\
5130 & 2004 Apr 9 & 19.41 & Y\\
6074 & 2004 Dec 16 & 19.84 & Y\\
6766 & 2006 Jun 6 & 19.70 & Y\\
\hline
\end{tabular}
\end{center}
\end{table}

%
%
%

\begin{figure}[htbp]
  \centering
   \includegraphics[width=0.45\textwidth]{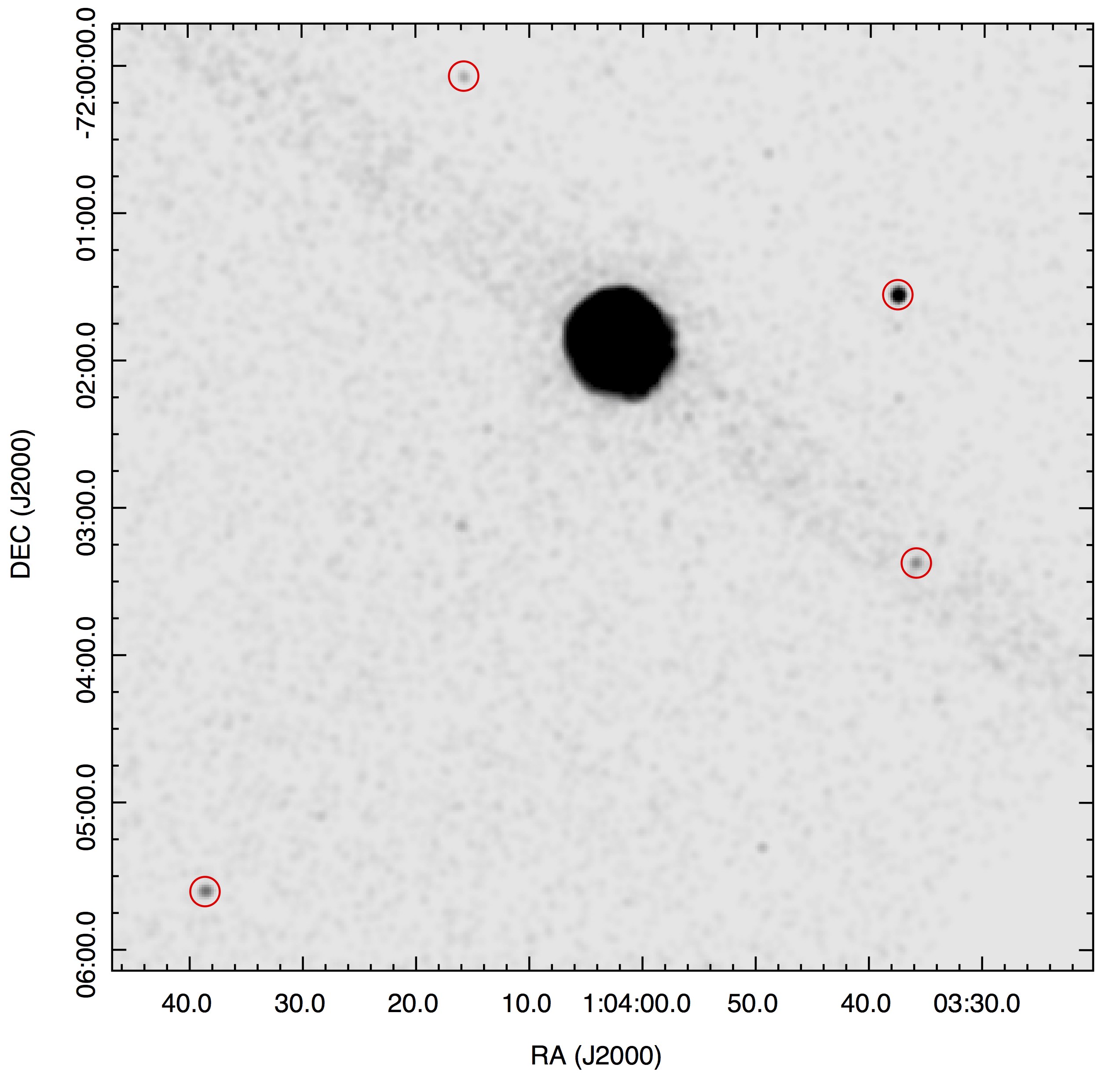} 
   \caption{Point sources used for estimating systematic registration uncertainty.  }
   \label{fig:register.point.sources}
\end{figure}

 To test the registration using point sources in fullframe observations to our method using the bright central feature, we compared the results for four fullframe observations registered relative to \texttt{ObsID} 1423 (see Table~\ref{tbl:3}) with both methods.
Figure~\ref{fig:register.point.sources} shows the locations of the point sources that were used for registration in relation to E0102.  There are four sources that are bright enough in all five fullframe observations to be used for registration. We registered \texttt{ObsID} 1423 to the 4 observations listed in Table~\ref{tbl:3}, and calculated the mean shift based on those 4 sources to register the images.  The mean shift derived from the point source registration and the shift derived from our method using the bright central feature are listed in Table~\ref{tab:4}.  The difference in the shift required to register the images using the two methods is slightly less than $0.1\arcsec$.  The mean difference between the two methods in the X direction is $0.03\arcsec$ and in the Y direction is $0.09\arcsec$.  This might indicate that the systematic uncertainty is larger in the Y direction since the bright central feature is elongated in this direction.  We adopt $0.10\arcsec$ as an estimate of the systematic uncertainty in our registration method. 

%

\begin{table}[htbp]
\centering
\caption{Comparison of shift results (in arcsec) using the Point Source (PS) and Central Knot registration methods}
\label{tab:4}
\begin{tabular}{l*{6}{c}r}
\hline\hline
Observation              & 5123 & 5130 & 6074 & 6766 \\
\hline
mean X PS shift         & -0.11 & -0.15 & -0.14 & -0.13 \\
mean Y PS shift         & \phantom{-}0.03 & -0.14 & \phantom{-}0.06 & -0.24 \\
registration X shift    & -0.11 & -0.14 & -0.11 & -0.19\\
registration Y shift    & \phantom{-}0.01 & -0.26 & \phantom{-}0.01 & -0.06\\
mean residual X & \phantom{-}0.03 & & &\\
mean residual Y & \phantom{-}0.09 & & &\\
%
%
\hline
\end{tabular}
\end{table}

\vfill
\subsection{Spectral Analysis}
\label{subsec:spectral.analysis}

We conducted a spectral analysis to convince ourselves that the regions we are using to measure the expansion of the blastwave or forward shock exhibit spectral properties consistent with swept-up interstellar material in the SMC  and to derive a temperature that can be used to infer a shock velocity.

We extracted spectra from five regions near the outer extent of the X-ray emission as shown in Figure~\ref{fig:regions.pileup.map} (top panel), which we believe to be representative of the forward shock emission. The regions have been adjusted to follow the curvature of the emission.
The widths of the regions are $1.0\arcsec$ to sample as small a region behind the forward shock as possible while maintaining sufficient statistics for spectral analysis.
We combined data from 36 \chandra observations to maximize the counts in the extracted spectrum (\texttt{ObsIDs}: 1308, 1311, 1530, 1531, 2843, 2844, 2850, 2851, 3520, 3544, 3545, 5123, 5124, 5130, 5131, 6042, 6043, 6074, 6075, 6758, 6759, 6765, 6766, 8365, 9694, 10654, 10655, 10656, 11957, 13093, 14258, 15467, 16589, 17380, 18418, 19850).  We included fullframe and subarray data in these extractions since these regions are negligibly affected by pileup as shown in Figure~\ref{fig:regions.pileup.map} (bottom panels).

\vfill
We fit the spectrum with a \texttt{vpshock} model in \texttt{XSPEC} (using the \texttt{wilm} abundances and the \texttt{vern} photoelectric constants) and allowed the abundances of O, Ne, Mg, and Fe to vary. We utilized a two-component model for the absorption, with one component for the Galactic line of sight absorption  (\texttt{tbabs}), $N_{H},_{Galactic}=5.36 \times 10^{20} cm^{-2}$ \citep{dickey1990}, and another with the SMC abundances specified in \cite{russell1992} (\texttt{tbvarabs}) and the best fitted value of $N_{H},_{SMC}=5.76 \times 10^{20} cm^{-2}$ determined from the \textit{XMM-Newton\xspace Reflection Gratings Spectrometer\xspace} presented in \cite{plucinsky2017}. Both absorption components were held fixed. 

A background model consisting of detector and sky background components was fit simultaneously with the source spectra.  The data were unbinned for the fitting process (the data have only been binned for display purposes) and the C statistic was used as the fit statistic.

\begin{figure}[htbp]
  \centering
   \includegraphics[width=0.45\textwidth]{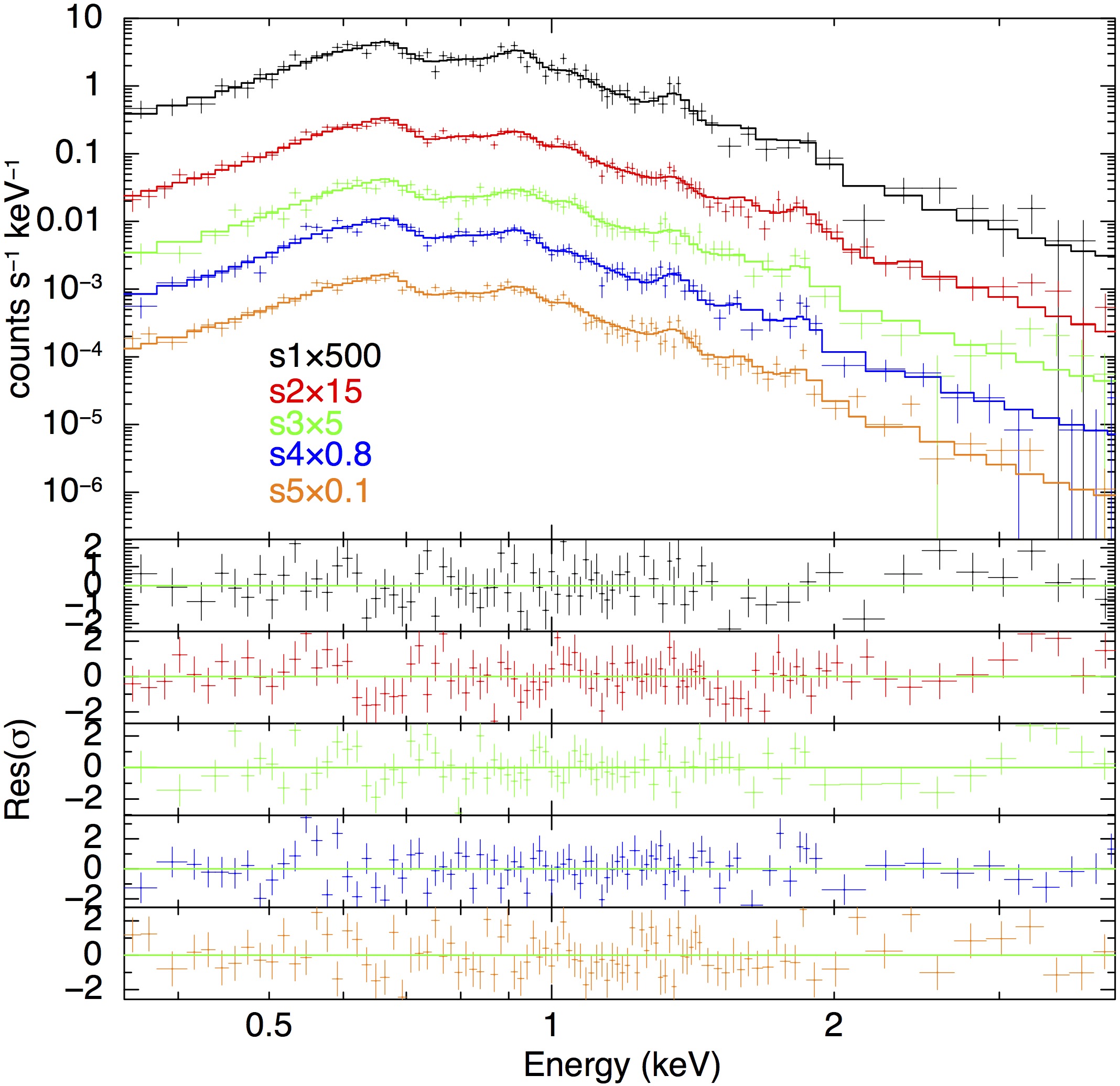}
   \caption{Spectrum of the blast wave regions shown in Figure~\ref{fig:regions.pileup.map} (top panel). The data are fit with a {\tt vpshock} model with parameters listed in Table~\ref{tab:blastwavespec}. The spectral data have been binned for display purposes only and an explicit background model and data (not shown) were fit simultaneously with the source model. } 
   \label{fig:specblast}
\end{figure}

\begin{figure}[htbp]
  \centering
   \includegraphics[width=0.45\textwidth]{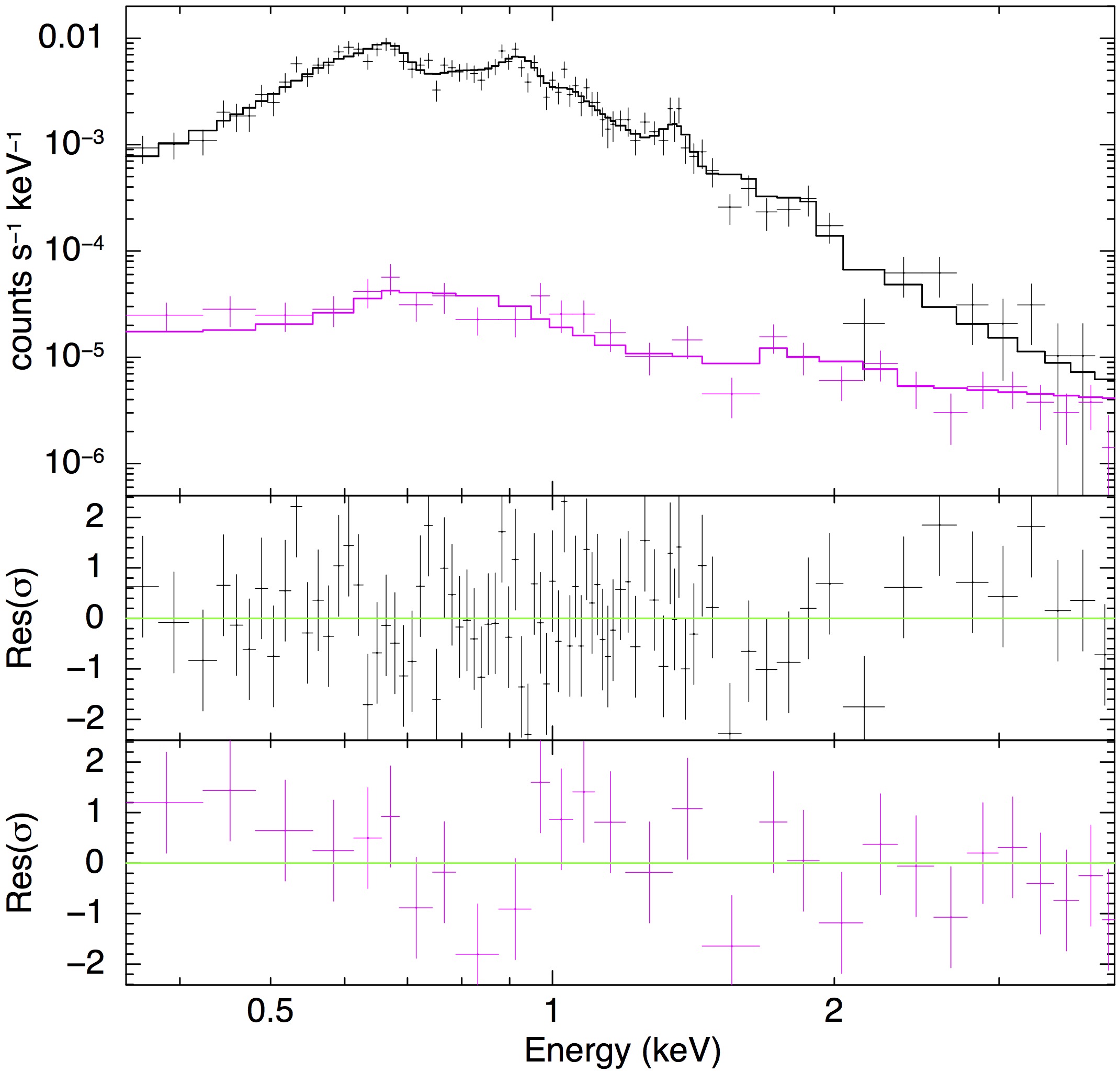}
   \caption{Spectrum of region s1 shown in Figure~\ref{fig:regions.pileup.map} (top panel). The data are fit with a {\tt vpshock} model with parameters listed in Table~\ref{tab:blastwavespec}. The spectral data have been binned for display purposes only and an explicit background model and data were fit simultaneously with the source model. The black data points and solid lines are for source data and model, respectively. The magenta data points and solid lines are for background data and model, respectively.} 
   \label{fig:specs1}
\end{figure}

The fit with a single {\tt vpshock} model produced an acceptable fit for all five regions with a C~statistic of 440 to 480 for 489 degrees of freedom (DOF). The fitted parameters are shown in Table~\ref{tab:blastwavespec} and the spectral fits with residuals are shown in Figure~\ref{fig:specblast}.
The temperatures for all 5 regions are consistent at the 1.0 sigma level.  The $n_{e}t$ are consistent at the 1.25 sigma level. The spectral fit of only region s1 is shown in Figure~\ref{fig:specs1}, to demonstrate that the source dominates over the background for most of the energy range considered here even for regions as faint as these forward shock regions.  The background region was selected to include emission on and off the transfer streak from E0102 itself.  The amount of transfer streak emission that affects our spectral extraction regions varies from observation to observation (from negligile to at most 6\%) as it depends on the roll angle of the observation on that date, which determines the orientation of the readout direction of the CCD on the sky.  Our background spectrum, and hence model, partially accounts for this relatively low level of contamination from the transfer streak.

The results show that the blastwave region has typical ISM abundances for the SMC ($\sim0.2-0.3$), although the Fe abundance is significantly lower than the expected SMC value.  The Ne abundance is higher than the expected SMC value.  This might indicate that the modeling has limited ability to distinguish Fe-L emission from Ne emission. These abundance values are much lower than the values in the bright ring which is dominated by ejecta emission from O, Ne, and Mg. \cite{sasaki2001} derives abundances of $4.7^{+4.6}_{-0.3}$, $7.1^{+6.4}_{-1.2}$, \& $3.0^{+0.3}_{-0.3}$ for O, Ne, \& Mg respectively for the entire remnant assuming a two component {\tt vgnei} model in {\tt XSPEC}.
We conclude that the regions from which we extracted a spectrum are consistent with
interstellar material heated by the blastwave.

\begin{table*}[htbp]
\caption{Spectral Model Parameters for the Blast Wave Regions. The model is $\mathrm{TBabs\times TBvarabs\times vpshock}$. The $\mathrm{TBabs}$ component represents the galactic absorption, $N_{H},_{Galactic}=5.36 \times 10^{20} cm^{-2}$, and the $\mathrm{TBvarabs}$ represents the SMC absorption of  $N_{H},_{SMC}=5.76 \times 10^{20} cm^{-2}$. The fitted elemental abundances of the $\mathrm{vpshock}$ component are listed in this table and the rest of the elemental abundances of the $\mathrm{vpshock}$ component are set to 0.2 solar abundances.}
\begin{center}
\label{tab:blastwavespec}
\begin{tabular}{ l c c c c c}
\hline\hline
Parameters & S1 & S2 & S3 & S4 & S5\\
\hline 
$kT_{e}\,\mathrm(keV)$ & $0.65^{+0.14}_{-0.07}$ & $0.75^{+0.06}_{-0.05}$ & $0.70^{+0.09}_{-0.11}$ & $0.71^{+0.10}_{-0.08}$ & $0.61^{+0.06}_{-0.04}$\\
Oxygen & $0.23^{+0.05}_{-0.04}$ & $0.27^{+0.05}_{-0.04}$ & $0.29^{+0.10}_{-0.06}$ & $0.27^{+0.04}_{-0.04}$ & $0.24^{+0.05}_{-0.04}$\\
Neon & $0.37^{+0.07}_{-0.05}$ & $0.29^{+0.04}_{-0.03}$ & $0.36^{+0.07}_{-0.06}$ & $0.36^{+0.06}_{-0.05}$ & $0.31^{+0.04}_{-0.04}$\\
Magnesium & $0.30^{+0.10}_{-0.08}$ & $0.13^{+0.04}_{-0.04}$ & $0.20^{+0.09}_{-0.07}$ & $0.30^{+0.09}_{-0.08}$ & $0.21^{+0.06}_{-0.05}$\\
Iron & $0.06^{+0.04}_{-0.02}$ & $0.06^{+0.02}_{-0.01}$ & $0.04^{+0.02}_{-0.02}$ & $0.10^{+0.03}_{-0.03}$ & $0.04^{+0.01}_{-0.01}$\\
$n_\mathit{e}\,t,(10^{11}\,\mathrm{cm}^{-3}\mathrm{s})$ & $1.50^{+0.99}_{-0.78}$ & $1.53^{+0.51}_{-0.39}$ & $2.32^{+2.83}_{-0.84}$ & $1.10^{+0.59}_{-0.37}$ & $2.24^{+0.93}_{-0.71}$\\
$Norm,(10^{-5})$ & $2.02^{+0.51}_{-0.60}$ & $4.76^{+0.57}_{-0.58}$ & $2.16^{+0.77}_{-0.40}$ & $5.57^{+0.57}_{-0.48}$ & $4.50^{+0.66}_{-0.70}$\\
\textit{C-statistic}\,(dof) & 440(489) & 477(489) & 454(489) & 480(489) & 470(489)\\
\textit{Pearson $\chi^2$}\,(dof) & 498(489) & 503(489) & 497(489) & 487(489) & 483(489)\\
goodness & 0.63 & 0.67 & 0.63 & 0.55 & 0.35\\
counts($\mathrm{10^{3}}$)& 1.67 & 2.77 & 1.54 & 2.21 & 2.82\\
area ($\mathrm{pixel^{2}}$)& 61.32 & 69.63 & 76.50 & 80.19 & 80.63\\
\hline
Weighted by $\mathrm{counts}$&&&&\\
$kT_{e}\,\mathrm(keV)$&&&$0.68^{+0.05}_{-0.05}$&&\\
$n_\mathit{e}\,t,(10^{11}\,\mathrm{cm}^{-3}\mathrm{s})$ &&&$1.73^{+0.46}_{-0.46}$ &&\\
\hline
\end{tabular}
\end{center}
\end{table*}

The fitted temperatures range from $0.61$~keV to $0.75$~keV, and the ionization time scales range  from $1.10\times10^{11}\;\mathrm{cm}^{-3}\;\mathrm{s}$ to $2.32\times10^{11}\;\mathrm{cm}^{-3}\;\mathrm{s}$. All the fits are statistically acceptable. We computed a counts-weighted average temperature of ${\mathrm{kT}_\mathit{e}}=0.68^{+0.05}_{-0.05}\;\mathrm{keV}$ and counts-weighted average ionization timescale of 
${n_\mathit{e}\,t} = 1.73^{+0.46}_{-0.46} \times 10^{11}\,\mathrm{cm}^{-3}\,\mathrm{s}$.  We will adopt these weighted values for the temperature and ionization timescale for calculations in \S\ref{subsec:e.i.temp.equil}.




\subsection{Radial Profile Analysis}
We extract 1-dimensional radial profiles from the model image generated from \texttt{ObsID}~1423 and the comparison observations, and fit them with a projected jump function to get the radius of the blast wave for each azimuthal direction.
We then calculate the expansion as the difference between the two radii. We fit the radial profiles with a theoretical model instead of shifting the radial profile of one epoch and comparing to the radial profile of the data from another epoch as was done by \cite{katsuda2008} for G266.2-1.2 and \cite{yamaguchi2016} for RCW~86 because the expected magnitude of the shift for E0102 is much smaller due to the fact that the distance to E0102 is 20-50 times larger than the distances to RCW~86 and G266.6-1.2. The expected expansion of the blast wave of E0102 is $0.35\arcsec$  for a 16 year baseline, which is comparable to the $0.246\arcsec$  bin size we have used for our radial profiles. For comparison, \cite{katsuda2008}   measure shifts as large as $6.0\arcsec$ for G266.2-1.2 and \cite{yamaguchi2016} measure shifts as large as $2.7\arcsec$ for RCW~86 which are several times larger than their bin sizes.  In addition, the E0102 data have significantly fewer counts per bin than the G266.2-1.2 and RCW~86 data which further motivates us to adopt the model-fitting approach.
\cite{Williams2018} adopted the approach of measuring the expansion of N103B in the LMC along four diameters rotated with respect to each other to form two orthogonal coordinate systems.  They computed the average of the expansion values on both sides of the diameter and adopted the average as their best estimate of the expansion since the values can be significantly different on either side and were in fact negative for one side of one diameter.  Using this method, \cite{Williams2018} find an average expansion of $0.31\arcsec$ over a 17.4~yr baseline which corresponds to a shock velocity of $4170~\mathrm{km}\,\mathrm{s}^{-1}$.

The expansion for each direction is calculated from:
\begin{equation} \label{eq:2}
f_{exp}=\frac{R_{2}-R_{1}}{R_{1}} ,
\end{equation}
where $R_{1}$ and $R_{2}$ are the radii of the remnant to the geometric center listed in Table~\ref{tab:geocenter}, from the model and later observation, respectively.  We adopt the estimate of the geometric center from Milisavljevic listed in Table~\ref{tab:geocenter}, see \S~\ref{sec:radii} for details.

\begin{figure}[htbp]
  \centering
   \includegraphics[width=0.45\textwidth]{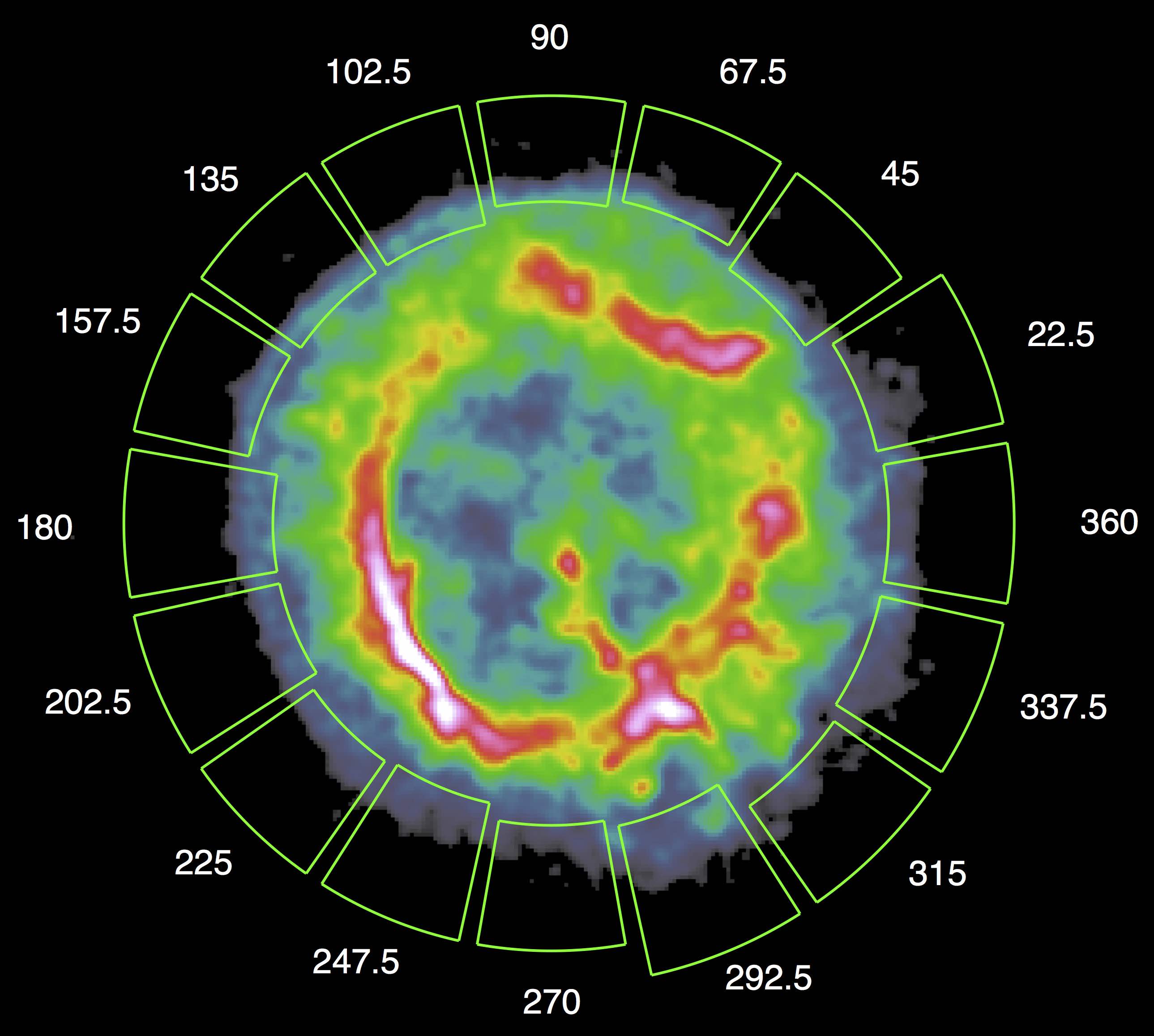} 
   \caption{The image from \texttt{ObsID} 1423 with the radial profile extraction regions overplotted. The radial profile extraction regions are indicated in green and the angular position are indicated in white. The widths of the regions vary depending on the structure of the remnant at that angular position. 
  }
   \label{fig:sectors.radial.profile}
\end{figure}

\begin{figure}[htbp]
  \centering
   \includegraphics[width=0.5
   \textwidth]{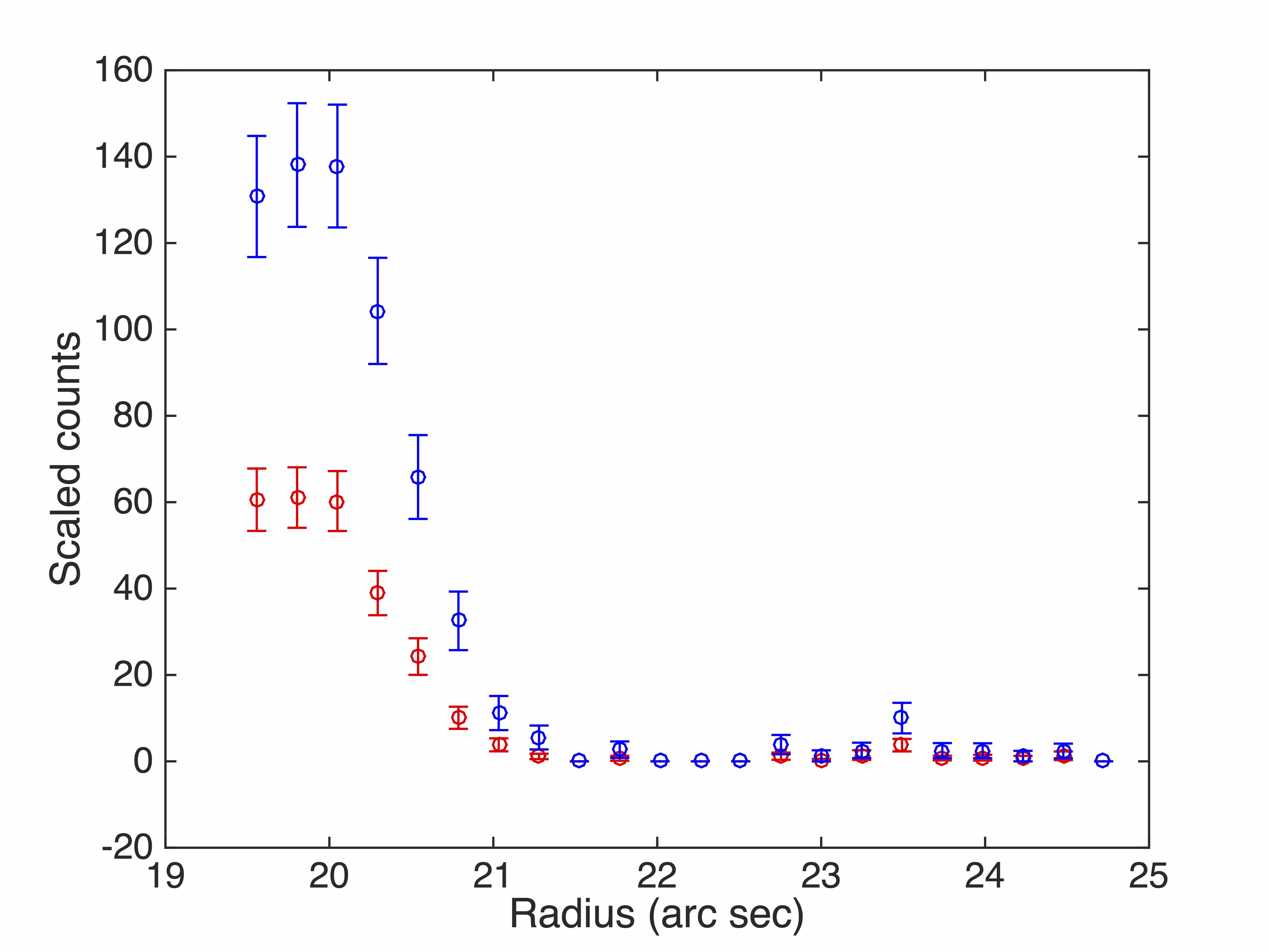} 
   \caption{Radial profile from \texttt{ObsID}~1423 with and without correction for the QE decrease.  The blue points are from \texttt{ObsID}~1423 itself.  The red points show the radial profile after correction for the loss of QE appropriate for the date of \texttt{ObsID}~17380.}
   \label{fig:qe_correction}
\end{figure}

The radial profiles are extracted from 16 azimuthal sectors, each covering $20^{\circ}$, and the position angles of these sectors start from $0^{\circ} (\equiv 360^{\circ})$ west, increasing counter-clockwise, stepping by $22.5^{\circ}$ increments to $337.5^{\circ}$ as shown in Figure~\ref{fig:sectors.radial.profile}. For each sector, the number counts in a radial bin at radius $r$ is obtained by the \ciao tool {\tt dmstat} summing the events within the region between radii of $r\pm 0.123\arcsec$. 
For the radial profiles from \texttt{ObsID} 1423, the counts in a bin are calculated by summing the events within the region and then weighting by the quantum efficiency loss and exposure time difference.  Figure~\ref{fig:qe_correction} shows the radial profile from \texttt{ObsID}~1423 for the azimuthal direction $\theta=135$.  The blue curve shows the radial profile for \texttt{ObsID}~1423 itself while the red curve shows the radial profile after it has been corrected for the decrease in QE appropriate for the date of \texttt{ObsID}~17380.
The area of an annular sector with a given radial width decreases as the annulus radius, $r$, decreases. The counts at $r$ are scaled with the ratio of the area of outermost region to the area of region at $r$. Note that the data for the radial profiles for the model image are not smoothed, smoothing was only applied to the model image used for registration. The C statistic was used for fits to the radial profiles as there are bins in the radial profile that have 0, 1, or 2 counts in the regions at larger radii than the forward shock; for plotting purposes, the \citet{gehrels1986} approximation ($\mathrm{error} = 1 + \sqrt{\mathrm{counts} + 0.75}$) was used as the uncertainty on the data points.  

\subsection{Radial Profile Model}
\begin{figure}[htbp]
  \centering
   \includegraphics[width=0.5\textwidth]{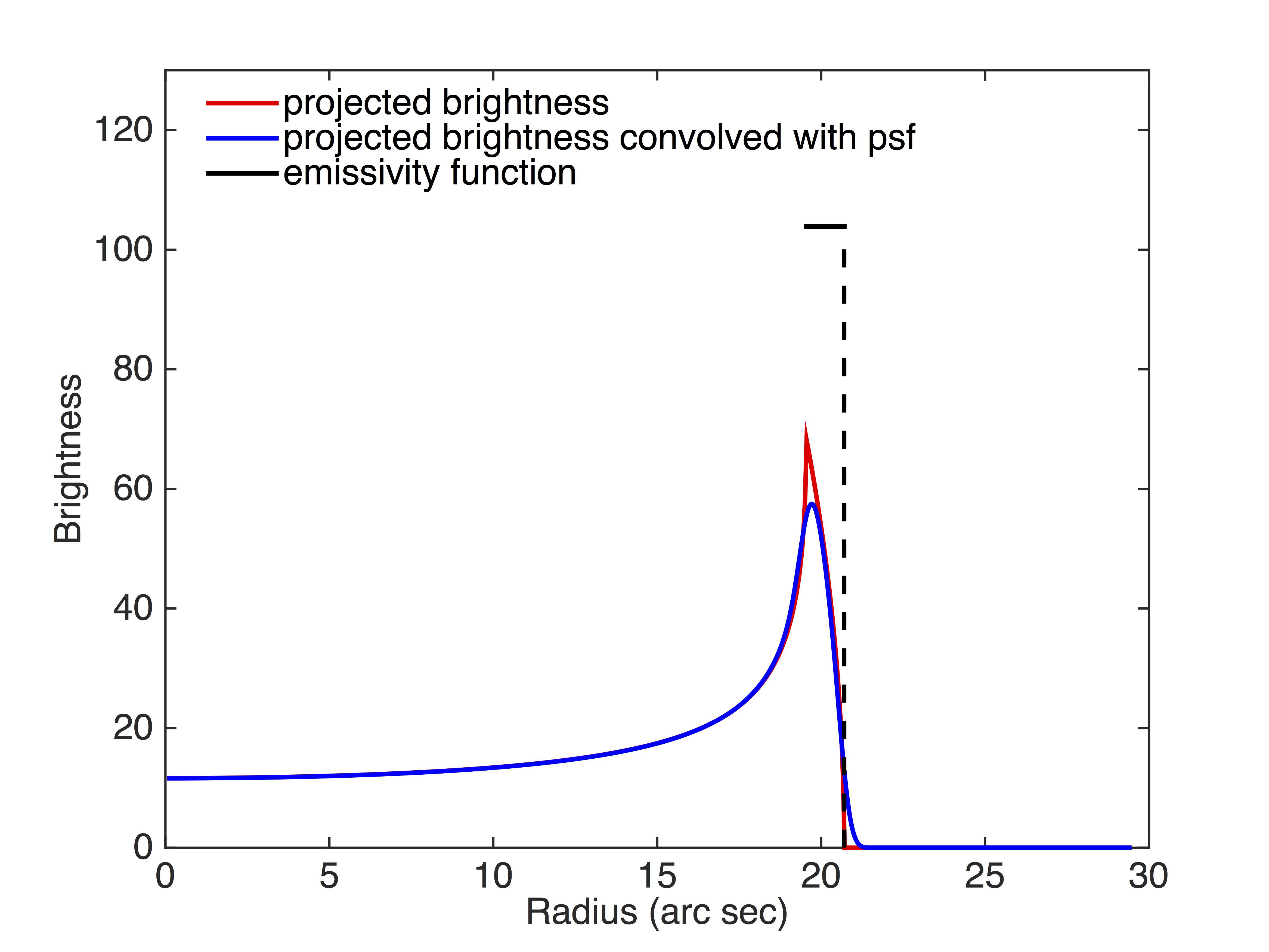} 
   \caption{Model for fitting the radial profiles.  The red curve gives the
   projection along the line-of-sight assuming the black emissivity function. The blue curve gives this distribution after it has been convolved with the \chandra PSF.}
      \label{fig:profile_model}
\end{figure}

To fit the radial profiles, we construct a model function which assumes a thin spherical shell of emitting material (in projection). The model profile is constructed by taking a slice through the thin spherical shell. These profiles are convolved with a 1D Gaussian function with $\sigma = 0.492\arcsec$ to account for the PSF of \chandra. Figure~\ref{fig:profile_model} displays the model after it has been convolved with the PSF. The emissivity function is assumed to be a step function with uniform intensity behind the blast wave, which is:
\[ e(r) =
  \begin{cases}
    I       & \quad \text{if }  R \geqslant r \geqslant R-d \\
    0       & \quad \text{if }   r > R\\
  \end{cases}
\]

\begin{figure}[htbp]
   \includegraphics[width=0.5\textwidth]{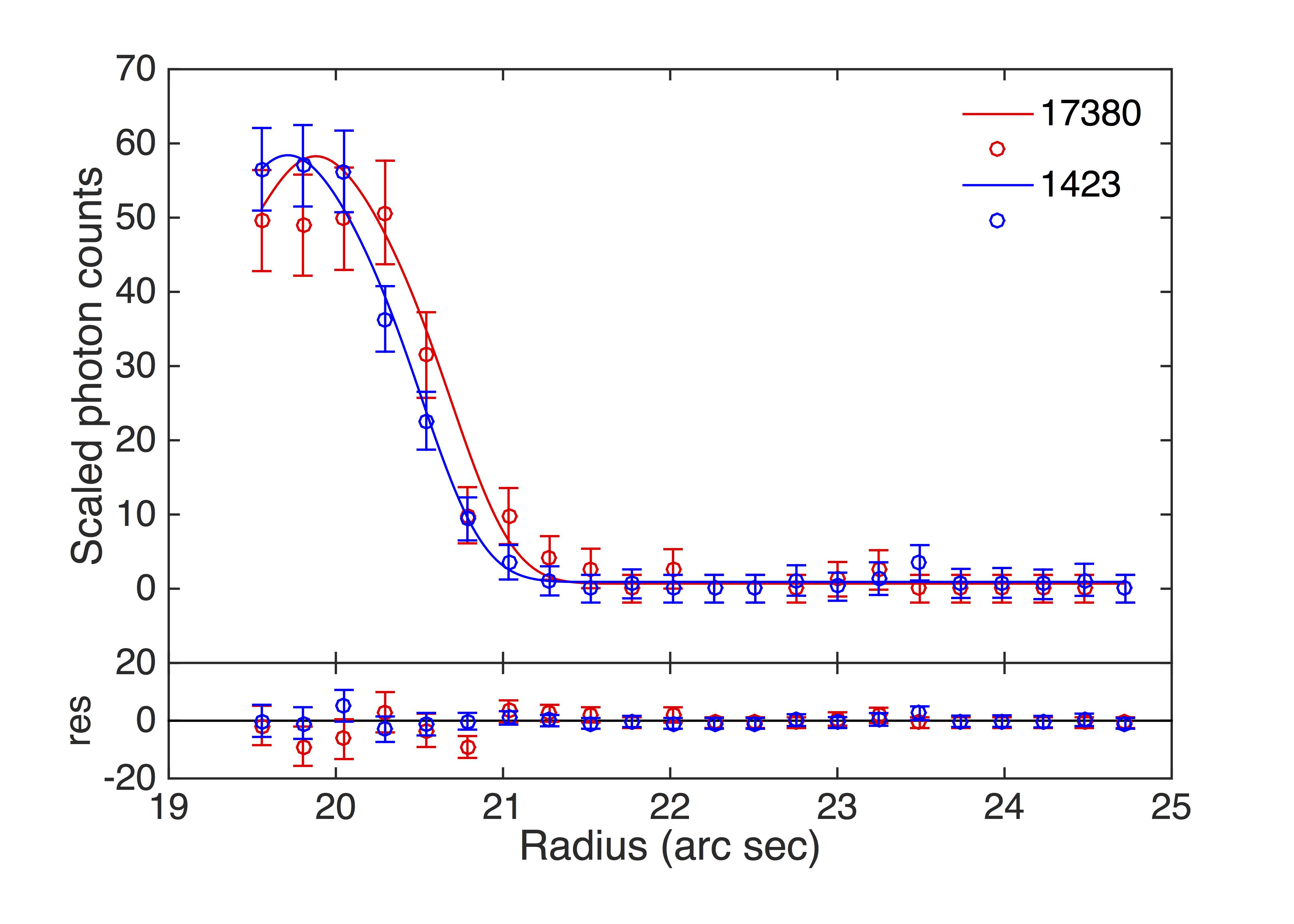} 
   \includegraphics[width=0.5\textwidth]{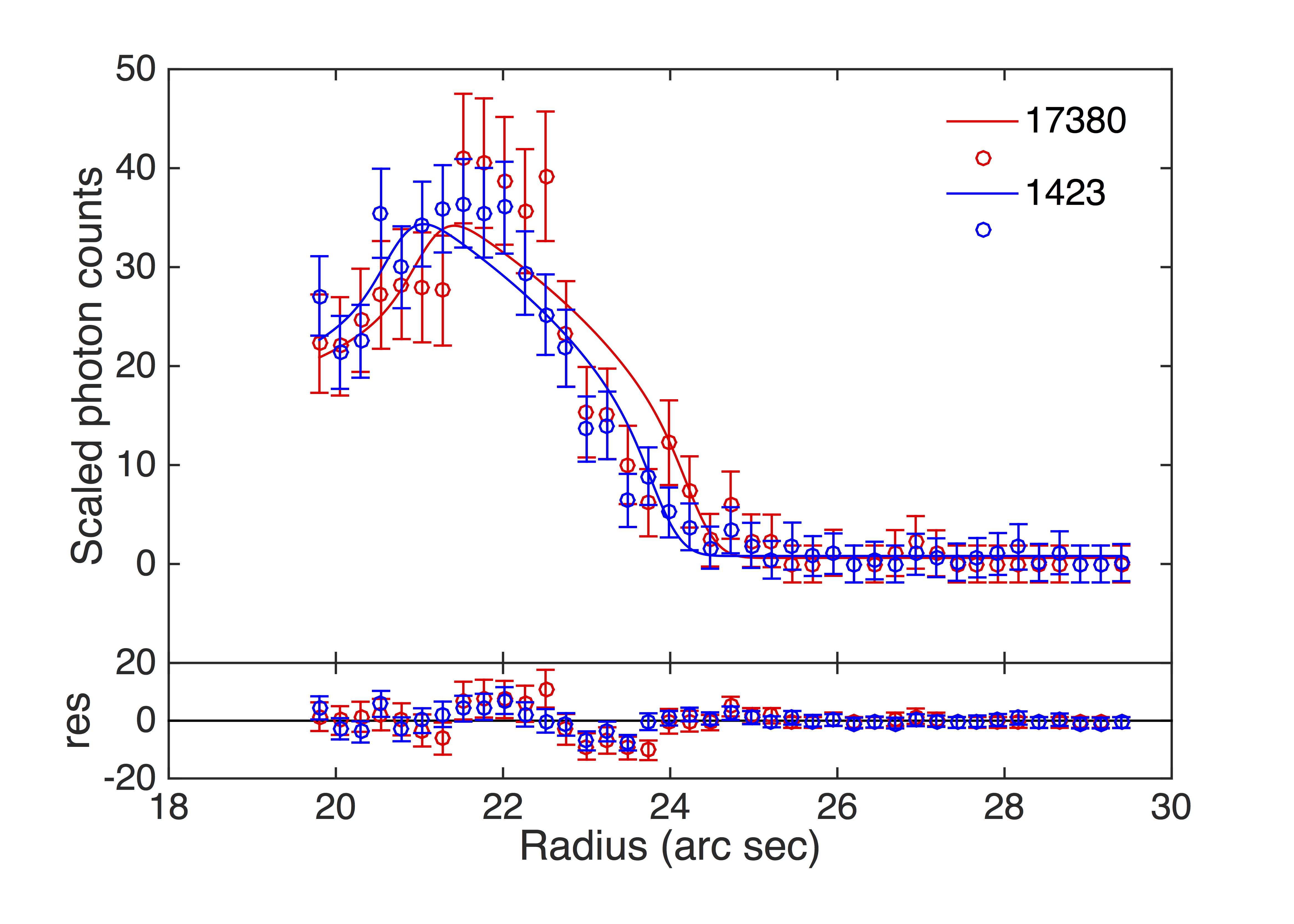}
   \includegraphics[width=0.5\textwidth]{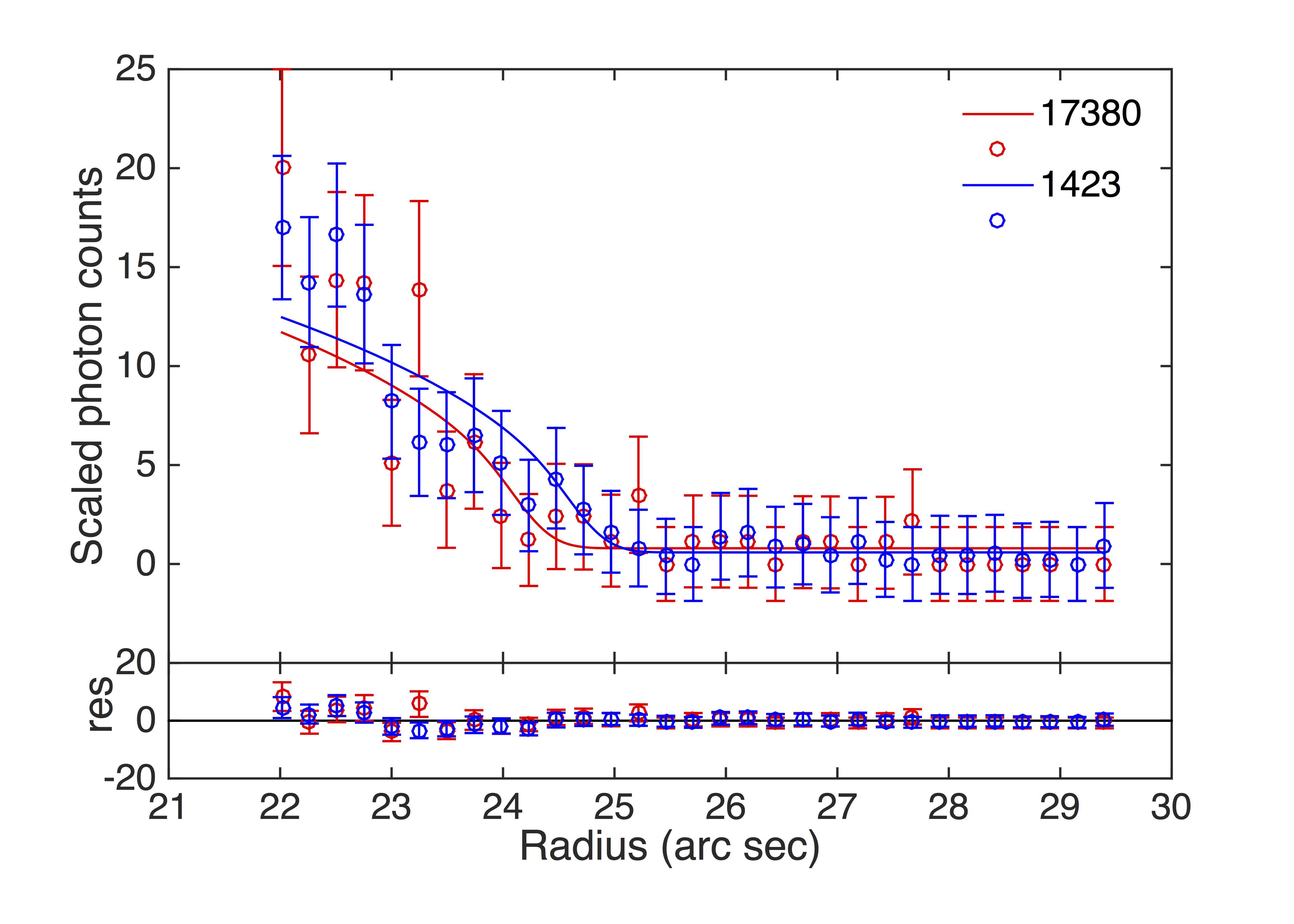}
   \caption{From top to bottom: radial profiles for azimuthal positions 135, 292.5, and 315 from \texttt{ObsIDs} 1423 and 17380.}
   \label{fig:radial.profile}
\end{figure}

Here, $I$ is intensity, $R$ is the radius of the blast wave, and $d$ is the width of blast wave. 
For fitting the 1D radial profiles, there are 4 free parameters, 3 of them from the shell emission model, $I$, $R$ and $d$, and the 4th is an additive uniform background level, $b$. We fit the profiles from \texttt{ObsID}~1423 first. Then the model with fixed $I$ and $d$, obtained from fitting profiles of \texttt{ObsID}~1423, is used to fit profiles from later observations, as these two values are not expected to change significantly with a time difference of up to 16 years.  The free parameters for the fits to the comparison data images are  $R$ and $b$. In this manner, we fit for the radii of the blast wave from 1423, $R_{1}$, and from later observations, $R_{2}$ in each of 16 directions.  Some sample fits are shown in Figure~\ref{fig:radial.profile} for three different angular positions from \texttt{ObsID}~17380 compared to \texttt{ObsID}~1423. The shape of the distributions for the radial profiles from \texttt{ObsID}~17380 are the same as those for \texttt{ObsID}~1423, the only parameters that change for the \texttt{ObsID}~17380 fits are $R_{2}$ and $b$.  The differences for the radii in the 16 directions are tabulated and used in the next step of the analysis. 

\subsubsection{Registration Bias Correction}

Our expansion results are sensitive to any remaining bias or systematic uncertainty (see \S\ref{subsubsec:reg.sys.uncertainty}) in the registration of the model image to the comparison data images.  We can estimate this bias by examining the radial differences computed in the previous section as a function position angle around the remnant.  Any bias in the registration would tend to increase/decrease the measured expansion in one direction while decreasing/increasing the expansion in the direction $180\arcdeg$ opposite.  Such a bias would manifest itself as a sinusoidal pattern in the radial differences as a function of position angle. An incorrect choice of the center for the annular sectors will also produce a sinusoidal variation, degenerate with the registration error. The uncertainties are large enough that it is impractical to extract azimuthal variation of the expansion, so we estimate an average expansion by zeroing out the sinusoidal component.
To correct for this bias, we fit the radial profile differences as a function of position angle with a model that has as parameters a shift in X and Y and the average expansion rate. 
For each test data (ObsID $i$) and for the reference (model) data we have radial profile differences as a function of position angle, $\theta_\mathit{j}$, $j=1 \cdots 16$.
For test data ObsID $i$ and reference data ObsID $i_0$, the profile difference for position angle $\theta_\mathit{j}$ is
\begin{equation}
\delta R_\mathit{j,i} = r_\mathit{i}(\theta_\mathit{j})
                      - r_\mathit{i_0}(\theta_\mathit{j})
\end{equation}
The sinusoidal fit function has the form:
\begin{equation}
\delta R_\mathit{j,i} = \Delta X_\mathit{i} \cos\theta_\mathit{j}
                      + \Delta Y_\mathit{i} \sin\theta_\mathit{j}
                      - \Delta R_\mathit{i}
 \label{eqn:sinusoid}                     
\end{equation}
where the $\delta R_\mathit{j,i}$ are the measured profile differences,
$\Delta X_\mathit{i}$ and $\Delta Y_\mathit{i}$ are the residual offset errors in the $X$ and $Y$ registration, and $\Delta R_\mathit{i}$
is the expansion estimate for observation $i$ relative to observation $i_0$.
The confidence intervals for the parameters are determined by the square roots of the diagonal entries of the covariance matrix, $\mathrm{cov}(\hat{\beta})$,
where $\hat{\beta}$ is the least-squares estimator for $\beta = (-\Delta X_\mathit{i}, -\Delta Y_\mathit{i},  \Delta R_\mathit{i})$.

\begin{figure}[htbp]
  \centering
   \includegraphics[width=0.45\textwidth]{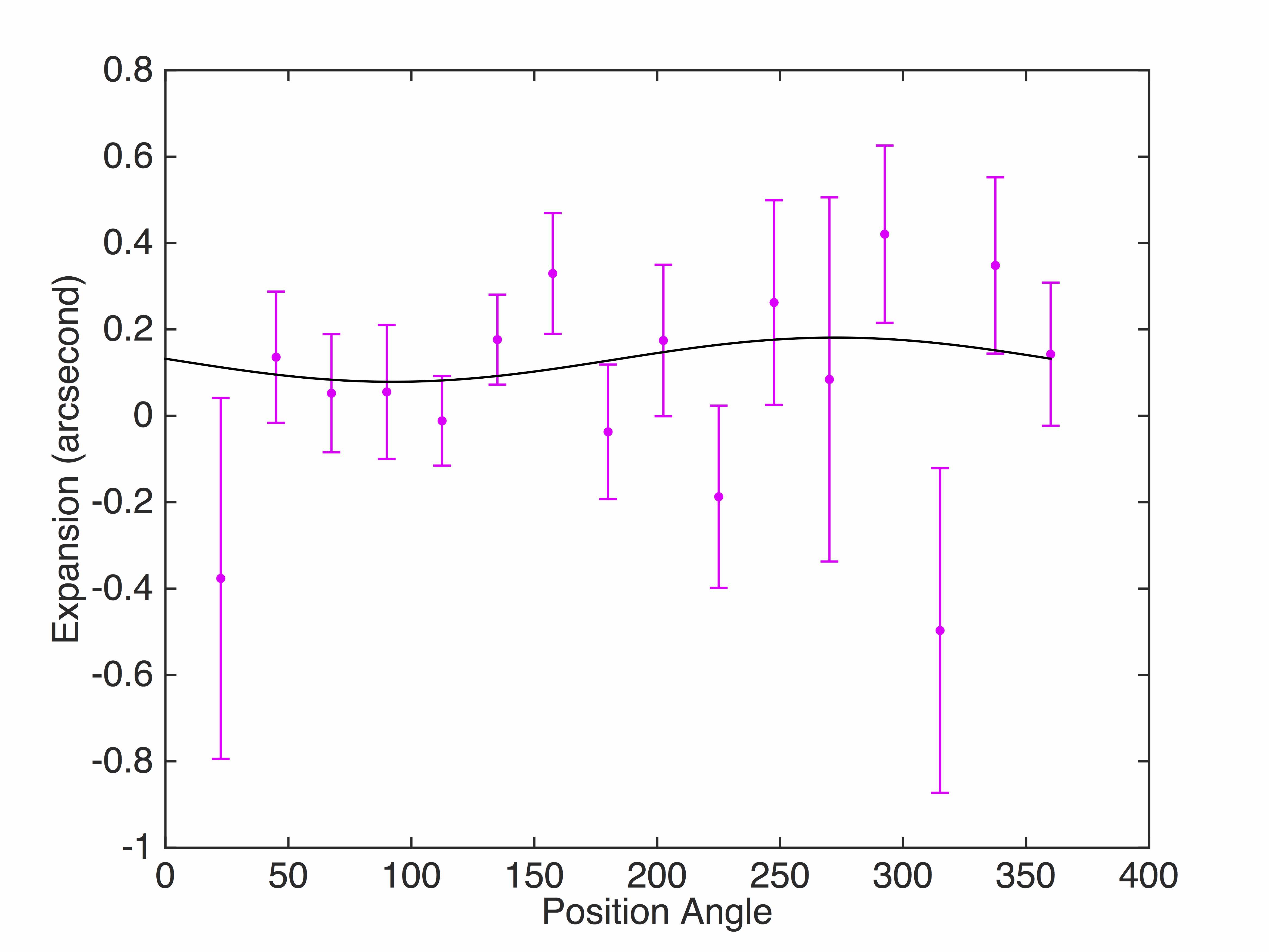} 
   \caption{Expansion measurements as a function of azimuthal position for \texttt{ObsID}~17380.  The black curve is the best fit for the model described in Equation~\ref{eqn:sinusoid}.
    }
    \label{fig:sinusoidal}
\end{figure}

With the expansion rates in 16 directions and an assumption of an approximately spherical geometry, we can obtain a global expansion rate for each pair of observations and the shift between the center of the two observations by fitting the expansion rates versus position angles with a sinusoidal function. 
By adopting this model, we are implicitly assuming that the expansion is the same in all 16 directions.  This clearly is not the case in many SNRS -- see, for example, \citet{vink2008} for the case of Kepler's SNR.  But it can be argued that the expansion of E0102 must be close to uniform around the remnant, based on the smooth, approximately symmetric outer extent of E0102. A sample fit for \texttt{ObsID}~17380 is shown in Figure~\ref{fig:sinusoidal}. The uncertainties on the data are determined from the uncertainties on the fitted radii from the radial profile fits.  It is clear that some directions have significantly larger uncertainties, such as $\theta=22.5,\;270.0,\;\mathrm{and}\;315.0$.  As is seen in Figure~\ref{fig:sectors.radial.profile}, these directions do not have a smooth, outer contour and hence the uncertainty on our radial profile fit is larger as our assumed model does not represent the data as well as it does in other directions.
The results of the fitted registration biases are listed in Table~\ref{tab:sinusoidal.pa}. The uncertainties for $\Delta X$, $\Delta Y$, $\Delta R$ are the 1.0~$\sigma$ statistical uncertainties from the sinusoidal fits.
As expected, the fitted shift values are small (less than $0.1\arcsec$ in all but two cases) and the average of the shift values is close to zero indicating a symmetric distribution of positive and negative shifts.
 The average of the magnitude of the shift in the X direction is $0.05\arcsec$ and in the Y direction is $0.09\arcsec$. This is consistent with our estimate of the systematic uncertainty in \S\ref{subsubsec:reg.sys.uncertainty}.

\begin{table*}
\caption{Registration Bias and Fitted Radial Shifts}
\begin{center}
\label{tab:sinusoidal.pa}
\begin{tabular}{ l c c c c c c}
\hline\hline
ObsID & $\Delta \mathrm{X}$ & $\sigma_{\Delta \mathrm{X}}$ & $\Delta \mathrm{Y}$ & $\sigma_{\Delta \mathrm{Y}}$ & $\Delta \mathrm{R}$ & $\sigma_{\Delta \mathrm{R}}$\\
&(arc sec)&(arc sec)&(arc sec)&(arc sec)&(arc sec)&(arc sec)\\
\hline 
3545 & 0.03 & 0.07 & 0.00 & 0.07 & 0.08 & 0.05\\
6765 & -0.07 & 0.06 & -0.16 & 0.07 & 0.04 & 0.05\\
6766\tablenotemark{a} & -0.01 & 0.06 & 0.04 & 0.06 & 0.04 & 0.04\\
9694 & -0.02 & 0.06 & -0.03 & 0.06 & 0.07 & 0.04\\
11957 & 0.08 & 0.06 & -0.17 & 0.06 & 0.02 & 0.04\\
13093 & 0.04 & 0.06 & -0.12 & 0.06 & 0.05 & 0.04\\
14258 & -0.04 & 0.06 & -0.07 & 0.06 & 0.01 & 0.04\\
15467 & -0.04 & 0.05 & 0.05 & 0.06 & 0.10 & 0.04\\
16589 & -0.15 & 0.07 & -0.08 & 0.08 & 0.08 & 0.06\\
17380 & -0.00 & 0.06 & 0.05 & 0.06 & 0.13 & 0.05\\
18418 & 0.06 & 0.06 & -0.21 & 0.07 & 0.09 & 0.05\\

mean shift x & -0.01 & mean shift y & -0.06 & & & \\
mean |x|  & 0.05 & mean |y| & 0.09 & & & \\
\hline
\end{tabular}

\begin{tablenotes}
        \footnotesize
        \tablenotetext{a} {full-frame observation}
\end{tablenotes}

              \end{center}
\end{table*}

\begin{table}[htbp]
\centering
\caption{Expansion Result and Shock Velocity for E0102.}
\label{tab:exp_results}
\begin{tabular}{cc}
\hline\hline
Expansion ($\%/{\rm yr}$)             & Shock velocity (km/s) \\
\hline
$0.025\pm0.006$       &  $1614\pm367$ \\
\hline
\end{tabular}
\end{table}

\begin{figure}[htbp]
 \centering
   \includegraphics[width=0.5\textwidth]{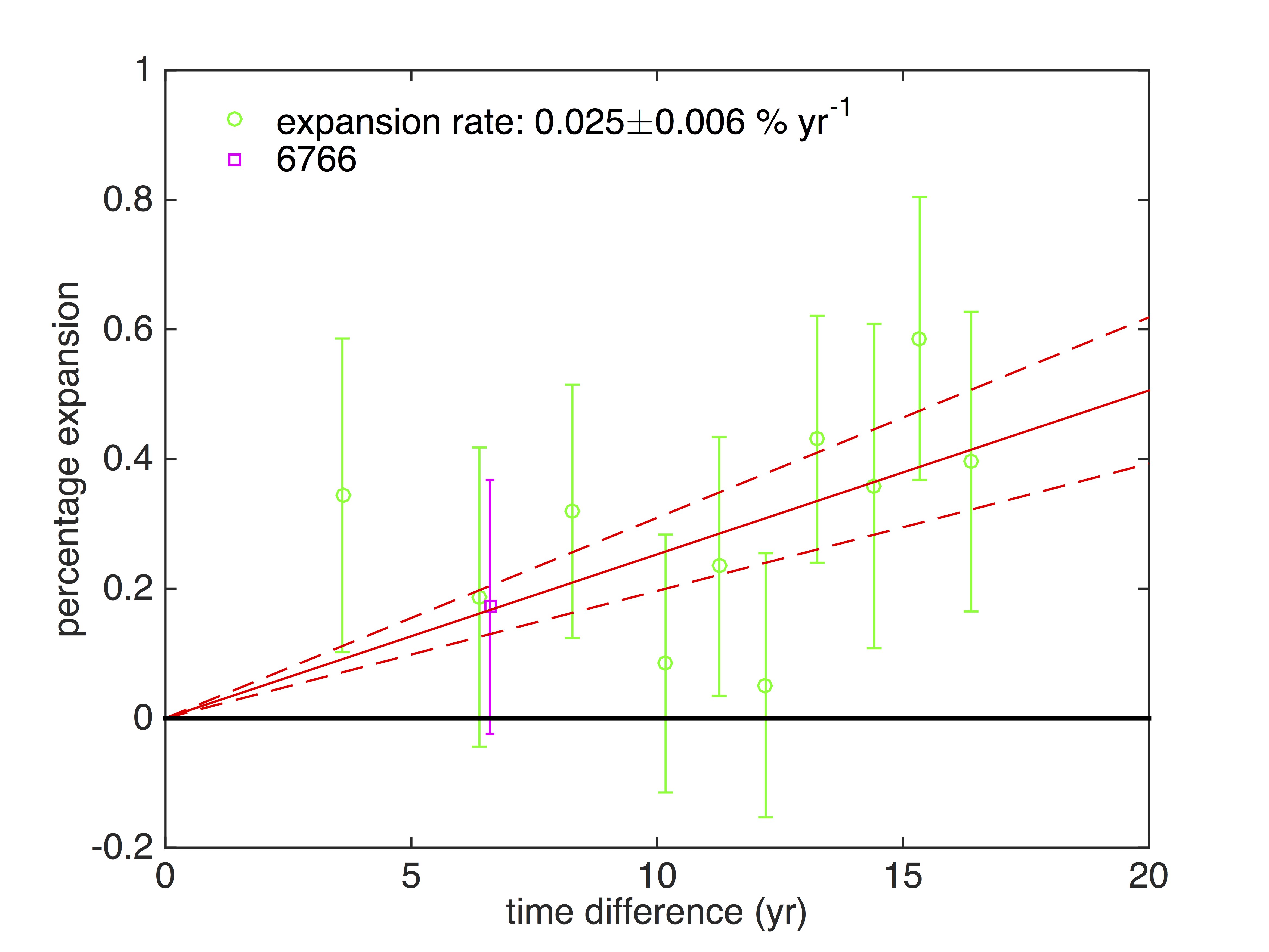} 
   \caption{The expansion measured in $\%$ versus time.  The green data points are the expansion rates measured by our method with the $1 \sigma$ statistical uncertainties. The magenta point is for \texttt{ObsID}~6766, the one observation for which we could register the images using point sources. The red line is the best fit line that is constrained to go through zero at ${\rm t=0}$ and the dashed red lines are
the $1 \sigma$ uncertainties on the slope. }

   \label{fig:expansion_rate_green}
\end{figure}

For each of the 11 comparison data images, the mean expansion rate and its uncertainty are listed in the last two columns of Table~\ref{tab:sinusoidal.pa}.  We include the latest full-frame observation, \texttt{ObsID}~6766, from 2006 in these results.
The evolution of the global expansion rate with time is shown in Figure~\ref{fig:expansion_rate_green}.  The data points are the measured expansions (percent) from the sinusoidal fits. The percent expansion versus time data were fit with a linear function, constrained to go through 0 at ${\rm t=0}$, to determine the best fitted value of the expansion rate.  It is worth noting that the magenta data point from \texttt{ObsID}~6766 is consistent with the adjacent data points and the best fitted line. This indicates that our registration method using the central bright knot produces consistent results when compared to registration with point sources.
Our measured expansion and the resulting shock velocity are listed in Table~\ref{tab:exp_results}. We derive a rate of $0.025\pm0.006\%~\rm{yr}^{-1}$ which is lower than the HRD00 result of $0.100\pm0.025\%~\rm{yr}^{-1}$. As described in the introduction, we are measuring the expansion of the forward shock while HRD00 measured the global expansion of the remnant.

\subsection{Determination of the Forward and Reverse Shock Radii}
\label{sec:radii}

\begin{table}[htbp]
\caption{Estimates of the Geometric Center}
\begin{center}
\label{tab:geocenter}
\begin{tabular}{l l l}
\hline\hline
Geometric centers & RA & DEC\\
\hline
\citet{finkelstein2006}        & 01h\,04m\,2.08s  & $-72^{\circ}\, 01\arcmin\, 52.5\arcsec$  \\
Milisavljevic$^\dagger$            & 01h\,04m\,2.354s & $-72^{\circ}\, 01\arcmin\, 53.23\arcsec$ \\
Ellipse of reverse shock & 01h\,04m\,2.048s & $-72^{\circ}\, 01\arcmin\, 52.75\arcsec$ \\
Ellipse of forward shock & 01h\,04m\,1.964s & $-72^{\circ}\, 01\arcmin\, 53.47\arcsec$ \\
\hline
\end{tabular}
\end{center}
\vskip-0.7em\quad$^\dagger$private communication\hfill
\end{table}

\begin{figure}[ht!]
  \centering
   \includegraphics[width=0.5\textwidth]
   {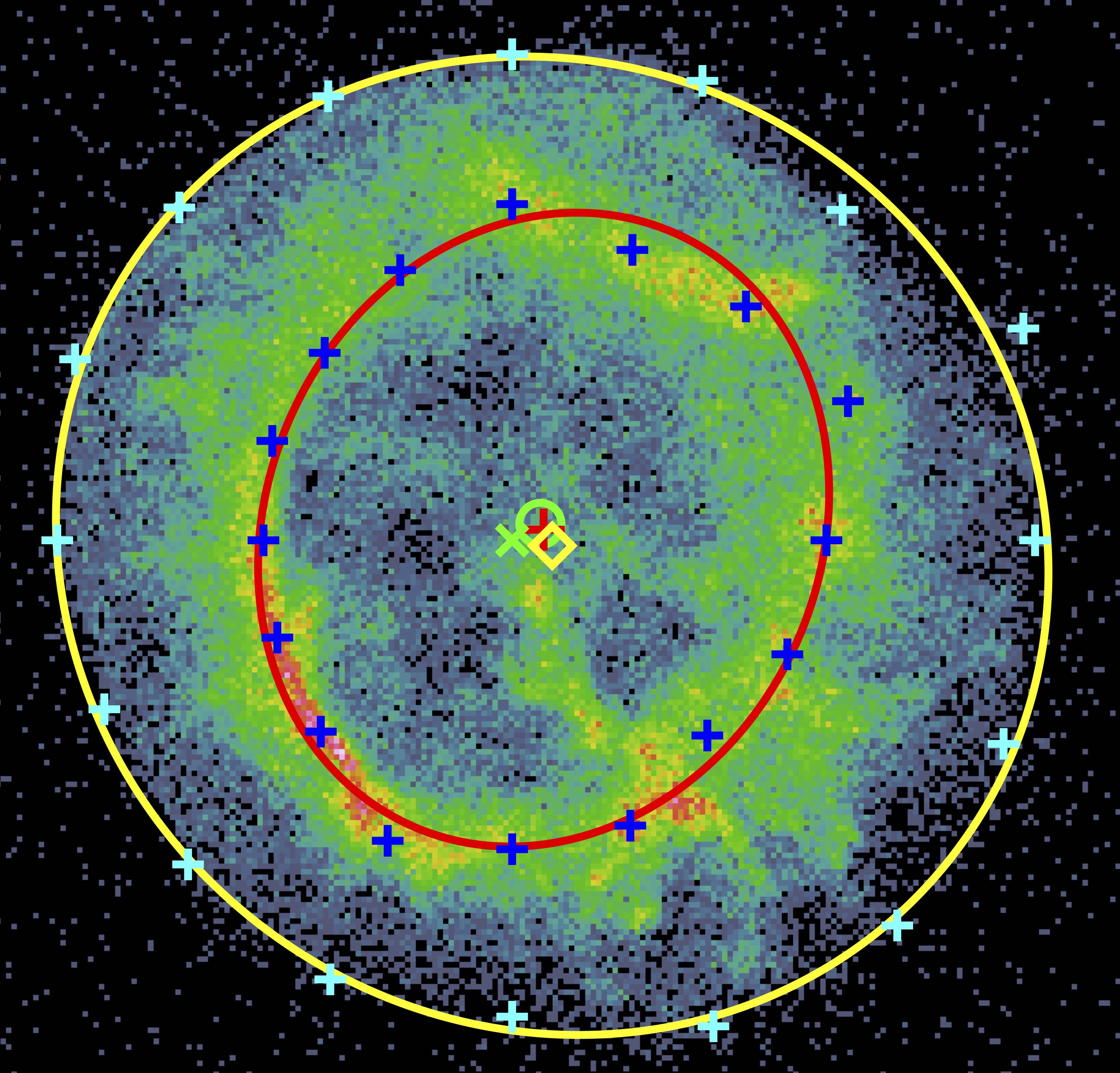}
   \caption{The ellipse fits for the forward and reverse shocks. The center positions are: Milisavljevic center (green $\times$), F06 center (green circle), fit based on X-ray radial profiles for the reverse shock (red crosses) and for the forward shock (yellow  diamond).  }
   \label{fig:ellipse}
\end{figure}

We estimate the position of the forward and reverse shocks using radial profiles extracted in the 16 directions described previously. For the forward shock position we adopted the value determined from our fits to the radial profile with the model that assumed a thin spherical shell geometry and blurring from the {\em Chandra}~PSF.  For the reverse shock we determined the point at which the emission increased to its peak value moving outward from our adopted center of the remnant.  In this manner, we derived 16 points each for the forward and reverse shock around the remnant.  If E0102 were perfectly spherically symmetric, we could fit the 16 points with a circle for both the forward and reverse shock to determine the radius and center.  However, the morphology does deviate from spherical symmetry as indicated by the larger extent of the forward shock in the southwest.  The morphology of the reverse shock is more complicated as indicated by the apparent elliptical shape and the bright feature in the northwest that is apparently concave compared to the generally convex shape of the rest of the ring.  The apparent elliptical morphology of the bright ring is consistent with the idea presented in  \cite{flanagan04} that the ejecta consist of two rings, one red-shifted and one blue-shifted, or a cylinder which is expanding. 
In this idea,
our line-of-sight is looking nearly down the major axis of the rings/cylinder.  Given this morphology, we fit the 16 points for the forward shock with an ellipse with the center, position angle, and semi-major and semi-minor axes free. We applied the same approach for the 16 points for the reverse shock. 
We compare the values of the ellipse centers from the fits to the forward and reverse shocks to the values of F06 and Milisavljevic~2017 (private communication) in Table~\ref{tab:geocenter} and in Figure~\ref{fig:ellipse}. There is excellent agreement between the centers derived from our fits and the centers derived from the optical expansion results.  
The fitted values for the semi-major, semi-minor axes, and rotation angle 
are displayed in Table~\ref{tab:ellipseerr}. We determine the radius of the forward shock to be $6.34 \pm 0.10 $ pc and the radius of the reverse shock to be $4.17 \pm 0.12$ pc.  For the forward shock we take the average of the semi-major and semi-minor axis radii (see Table~\ref{tab:ellipseerr}) with the range of values providing an estimate of the systematic uncertainty.  For the reverse shock, we adopt the semi-major axis for the radius, appealing again to the concept of overlapping rings and/or a cylindrical geometry seen somewhat off-axis, and adopt the statistical uncertainty on the fit as the uncertainty.  If we assume the X-ray bright ring is circular in nature and appears elliptical due to our off-axis viewing angle, we can estimate that angle from our measured ellipticity.   From the minor and major axis values in Table~\ref{tab:ellipseerr}, we estimate an angle of ${31^{\circ}}^{+6^{\circ}}_{-7^{\circ}}$.

\begin{table}[htbp]
\caption{Fitted Values for the Ellipses for the Forward and Reverse Shocks}
\begin{center}
\label{tab:ellipseerr}
\begin{tabular}{l c c c}
\hline\hline
Parameters & Unit & Forward shock & Reverse shock\\
\hline
xc error & arcsec & 0.074 & 0.023\\
yc error& arcsec & 0.075& 0.072\\
Major Axis & pc & $6.52^{+0.12}_{-0.12}$ & $4.17^{+0.12}_{-0.12}$\\
Minor Axis & pc & $6.16^{+0.16}_{-0.12}$ & $3.57^{+0.12}_{-0.13}$\\ 
Rotation Angle& degree& $-37.61^{+2.67}_{-2.69}$ & $65.85^{+1.16}_{-1.16}$\\
\hline
Radius  & pc & $6.34^{+0.10}_{-0.10}$ & $4.17^{+0.12}_{-0.12}$  \\
\hline
\end{tabular}
\end{center}
\end{table}

\section{One Dimensional Shock Models}
\label{sec:one.dimensional.shock.models}

We generate analytic solutions following the work of \citet{truelove99,truelove99erratum}, as extended by \citet{laming2003}, \citet{hwang2012} and \citet{micel2016}.
\citeauthor{truelove99} noted that the Euler equations do not contain any dimensioned parameters,
and that dimensional aspects are introduced through the initial conditions. They
introduced ``unified solutions'', single dimensionless solutions merging
similarity solutions for the initial ejecta dominated stage
\citep{chevalier1982}
through to the asymptotic Sedov-Taylor similarity solution \citep{sedov1959,taylor1950}. Three dimensioned parameters are introduced, $E$,
the energy
of the ejecta (explosion energy), $M_\mathit{ej}$, the mass of the ejecta, and $\rho_0$, the
mass density of the preshock medium

The models assume an initial ejecta distribution expanding homologously
with $v \propto r$ up to a maximum velocity $v_\mathit{ej}$ at the edge of the ejecta. The ejecta profile is given as
\begin{equation}
\rho_\mathit{ej}(r,t) = \rho_\mathit{ej}\left(v,t\right) \equiv \frac{M_{ej}}  {\left( v_\mathit{ej} \,t\right )^{3}} f\left(\frac{v}{v_\mathit{ej}}\right)
\label{eqn:ejecta.profile}
\end{equation}
where the dimensionless structure function, $f$ is
\begin{equation}
f(\omega) =
  \begin{cases}
    f_{0}             & 0 \leq \omega \leq \omega_\mathit{core} \\
    f_{n}\omega^{-n}  & \omega_\mathit{core} \leq \omega \leq 1
  \end{cases}
\label{eqn:ejecta.structure.function}
\end{equation}
with $\omega$ defined as
\begin{equation}
\omega \equiv \frac{v}{v_\mathit{ej}}
\label{eqn:def.omega}
\end{equation}
%

For steep power laws (large $n$) a constant density ``core''
is introduced. The core is taken to have constant density for simplicity; all that is required is that the core have a sufficiently shallow power law distribution. \citeauthor{truelove99} note that the distribution of mass and energy depends on $n$; for
$n < 3$, mass and energy are concentrated in the outer (high speed) ejecta,
while for $n > 5$, mass and energy are concentrated in the inner (low speed)
ejecta.  Based on \citet{matzner1999ApJ}, it is expected that $n$ is relatively
large, $n \gtrsim 7$. Our explorations showed that the results were relatively insensitive to $n$ for $n > 7$; in our treatment below we set $n = 9$.

The preshock ambient medium is also assumed to follow a power law distribution
\begin{equation}
\rho \propto \rho_0 r^{-s}
\label{eqn:ambient.medium.profile}
\end{equation}
where $s$ is 0 or 2, and $\rho_0$ is the ambient medium mass density just ahead of the shock. For the $s=0$ case (constant density ambient medium) the value of $\rho_0$ is constant. For the $s=2$ case (ambient density falls as $1/r^2$, appropriate for a constant stellar wind), the value of $\rho_0$ pertains to a particular choice for the blast wave radius, $R_\mathrm{b,0}$, so that $\rho(r) = \rho_0/(r/R_\mathrm{b,0})^2$.
The introduction of a power law ambient
medium density distribution does not introduce additional dimensioned constants
and the ejecta structure function is a dimensionless function,
so asymptotic similarity solutions can still be constructed. Note that the $r^{-2}$ ambient medium is assumed to be stationary ($v_\mathit{wind} \equiv 0$). This implies that the velocity of the constant wind should be much smaller than the forward shock velocity. Incorporating a significant wind velocity would introduce an additional dimensioned parameter, significantly complicating the solution.

\cite{truelove99} produced a detailed treatment of the $s=0$ case, but only a limited treatment
of the $s=2$ case. \cite{hwang2003}, \cite{laming2003}, and \cite{micel2016}
extend the treatment to a more detailed consideration of the $s=2$ case. We
refer the reader to these papers for the details. Here, we touch mainly on
those aspects relevant to our analysis. In particular, we consider
``swept-up'' mass, and ``unshocked ejecta'', subject to the caveat that
the models are highly idealized, and that the actual situation is far more complex.

For swept up mass in the $s=2$ case, this is presumably the stellar wind
of the progenitor, so swept-up mass plus ejecta mass plus $\sim 1.5\,M_\sun$ for
a compact remnant would be a lower limit to the progenitor mass, since the blast wave radius may not have reached the edge of the stellar wind. We note that HST imaging \citep[WFPC2 and ACS;][]{finkelstein2006} and \cite{gaetz2000} show a partial ``bowl'' of 
[\ion{O}{3}] emission, mostly complete southeast through west; if the [\ion{O}{3}] emission is from the photoionized wind of the progenitor,
the stellar wind likely extends no more than $\sim 1.5-2$ times the current blast wave radius.
Note that for a $r^{-2}$ wind, the swept-up mass is proportional to the blast wave radius, so the above implies a factor of $\sim 2$ uncertainty in swept-up mass.
In the case
of $s=0$, such a simplistic interpretation is not available. The ``constant density'' preshock medium could be an average over some combination of
progenitor mass loss and ambient medium.

Strong caveats are also necessary for the treatment of unshocked ejecta.
We use this estimate as a sanity check, and for obtaining a rough idea of
how much (unshocked) Fe might have been produced, since there is no definitive detection
of substantial amounts of X-ray emitting Fe in E0102. 
The ``unshocked'' ejecta would also include ejecta emitting optically in [\ion{O}{3}] and [\ion{O}{2}] \citep{tuohy1983} which have not yet been shocked to X-ray emitting temperatures.

\subsection{Remnant Evolution}
\label{subsec:remnant.evolution}

To describe the evolution of the blast wave and of the reverse shock 
$R_\mathit{b,0}$, so that $\rho(r) = \rho_0(r/R_\mathit{b,0})^{-s}$
 for the $s=2$ case we follow \cite{micel2016},
while for the $s=0$ case, we follow \cite{truelove99}. 
The $s=2$ (constant wind) case is an oversimplification of the evolution of the star before the explosion; however, it is useful to contrast this with the uniform medium case.

Our models  depend on three dimensioned parameters, $E_{51}$ (the explosion energy in units of $10^{51}\,\mathrm{erg}$), $M_\mathit{ej}$ (the ejecta mass in units of \msol), and $\rho_0$ (the preshock ambient medium mass density), plus two dimensionless parameters ($n$, $s$), for a total of five adjustable parameters.
The ambient medium mass density is
related to the H number density $n_0$ by $n_0 = (\rho_0/\mu_\mathrm{H})/(1\,\mathrm{cm}^{-3})$ where $\mu_\mathrm{H}$ is the mean mass per hydrogen nucleus, assuming cosmic abundances.
As noted in \S\ref{sec:one.dimensional.shock.models}, the solutions were relatively insensitive to $n$ for $n\ge 7$, and we adopt $n=9$. We consider the $s=0$ and $s=2$ cases separately. For each $s$ value, we 
consider ejecta masses of 2, 3, 4, 5, and 6\,\msol. Our observational constraints are the measured blast wave velocity, $v_\mathit{b}$,
the observed blast wave radius, $R_\mathit{b}$, and the reverse shock radius, $R_\mathit{r}$ \citep[based on the distance to E0102, taken to be 60.6\,kpc,][]{hilditch2005}. In each case, the other two free parameters, $E_{51}$ and $\rho_0$, are varied until
until the blast wave velocity $v_\mathit{b}$,
blast wave radius $R_\mathit{b}$, and reverse shock radius $R_\mathit{r}$ are matched.

Once the match is found, some derived quantities can be
evaluated based on the parameters of the model: ejecta ``core'' mass, remnant age, reverse shock velocity $v_\mathit{r}$, swept-up mass, and
unshocked ejecta mass.
We estimate the swept-up mass by integrating the initial ambient medium profile out to the current blast wave radius, $R_\mathit{b}$. The unshocked ejecta mass is obtained from Eqs.~\ref{eqn:ejecta.profile} \& \ref{eqn:ejecta.structure.function} by integrating from $v_\mathit{ej}$ down to the reverse shock velocity $v_\mathit{r}$ making use of Eq.~\ref{eqn:def.omega}. 

A deceleration parameter, $m$, for the blast wave, is also of interest. The blast wave radius $R_\mathit{b}$ varies with time as
\begin{equation}
R_\mathit{b} \propto t^\mathit{m}
\end{equation}
where $m$ varies from $m=1$ (initial free expansion of the ejecta), to $m=2/5$ asymptotically as the Sedov-Taylor similarity solution is reached.
The resulting values are given in Table~\ref{tab:model.results}.

\begin{table*}[!htbp]
\caption{Models for ejecta profile $n=9$, $R_{b}=6.34$~pc, $R_{r}=4.17$~pc and $v_{b}= 1614 \mathrm{km~s}^{-1}$}
\begin{center}
\label{tab:model.results}
\begin{tabular}{ l c c c c c c c c c c c c c}
\hline\hline
Parameters & Symbol(units) &&&$s=2$&&&&&$s=0$&&\\
\hline
  Ejecta mass
& $M_\mathit{ej}(\msol)$
& 2
& 3
& 4
& 5
& 6
& 2
& 3
& 4
& 5
& 6 \\
  Explosion energy
& $E(10^{51}\,\mathrm{erg})$
& 0.34
& 0.51
& 0.68
& 0.85
& 1.02
& 0.87
& 1.31
& 1.74
& 2.18
& 2.61 \\
  Circumstellar density
& $\rho_{0}(\mathrm{amu~cm}^{-3})$
& 0.22
& 0.33
& 0.44
& 0.55
& 0.66
& 0.85
& 1.27
& 1.69
& 2.11
& 2.54 \\
  Swept-up mass &
$\msol$
& 17.4 
& 26.1
& 34.8
& 43.5
& 52.2
& 22.1
& 33.2
& 44.2
& 55.3
& 66.3 \\
  core mass
& $\msol$
& 1.3
& 2.0
& 2.7
& 3.3
& 4.0
& 1.3
& 2.0
& 2.7
& 3.3
& 4.0 \\
Unshocked mass
& $\msol$ 
& 0.06
& 0.09
& 0.12
& 0.15
& 0.18
& 0.05
& 0.08
& 0.10
& 0.13
& 0.16 \\
\hline
\end{tabular}
\end{center}
\end{table*}

\begin{table}[htbp]
\caption{The age, reverse shock velocity, upstream ejecta velocity, downstream ejecta velocity, and expansion parameter for the s=2 and s=0 cases.}
\begin{center}
\label{tab:model.error}
\begin{tabular}{ l c c c}
\hline\hline
$\mathrm{Parameters}$ & $\mathrm{Symbol(units)}$ & $s=2$ & $s=0$\\
\hline 
Age & $\mathrm{yr}$ & $2642^{+1002}_{-489}$ & $1730^{+552}_{-400}$\\
Reverse shock velocity & $\mathrm{v_{r} (km s^{-1})}$ & $923^{+413}_{-314}$ & $37^{+306}_{-185}$\\
Upstream ejecta velocity & $\mathrm{v_{r,u} (km s^{-1})}$ & $1544^{+369}_{-402}$ & $2359^{+621}_{-597}$\\
Downstream ejecta velocity & $\mathrm{v_{r,d} (km s^{-1})}$ & $1078^{+387}_{-333}$ & $618^{+384}_{-288}$\\
Expansion parameter & $\mathrm{m}$ & $0.69^{+0.06}_{-0.01}$ & $0.45^{+0.01}_{-0.01}$\\
\hline
\end{tabular}
\end{center}
\end{table}

\subsection{Model Results}

The range of values in Table~\ref{tab:model.results} may now be compared to theoretical predictions, measurements of other SNRs, and measurements at other wavelengths. The explosion energy of a single star is expected to range between $0.5-1.8\times10^{51}$~ergs depending on the mass, metallicity and rotation of the progenitor \citep[see][and references therein]{sukhbold2016}. The ejecta masses of the X-ray emitting O and Ne (the dominant constituents in E0102) have been estimated to be $\sim5.7$ and $\sim2.2\,\msol$ respectively by \cite{flanagan04}. However, they note that their assumption of a pure metal plasma results in the largest estimate of the ejecta mass and any hydrogen in the plasma would only reduce their ejecta estimates. 
This can be compared to estimates of $2-4\,\msol$ for Cas~A \citep{hwang2012}. In addition to the ejecta observed in X-rays, there is also a contribution from the ejecta observed in the optical. A star with an initial mass of $25-40\,\msol$ is considered to be capable of producing these large ejecta masses of O and Ne \citep[see][and references therein]{nomoto1997, sukhbold2016}.
The masses of the most massive stars observed in the SMC have a maximum value of $\sim90\,\msol$ \citep[][]{massey2000}. 

We are able to estimate the density of the preshock medium from our spectral fits described in \S\ref{subsec:spectral.analysis}.  Adopting the normalizations in Table~\ref{tab:blastwavespec} 
to calculate the emission measure and assuming emission from a thin shell geometry, we derive an H number density of $n_{0}=0.86\pm \mathrm{0.03~cm}^{-3}$.
There is at least a factor of two uncertainty on this calculation in addition to the statistical uncertainty due to our assumptions of the geometry and filling factor.


The estimated age and reverse shock velocity for the s=0 and s=2 cases are listed in Table~\ref{tab:model.error}.
The results for the $s=2$ case indicate an age of $\sim2600\;\mathrm{yr}$, a reverse shock velocity of $\sim900~\mathrm{km}\,\mathrm{s}^{-1}$, and an expansion parameter of $\sim0.7$.  The age is larger than those estimated by F06 ($2122\pm644$~yr and $2054\pm584$~yr) but within the uncertainties. The reverse shock velocity and the expansion parameter indicate that the remnant has not yet reached the Sedov phase of evolution. The results for the $s=0$ case indicate an age of $\sim1700$~yr, a reverse shock velocity close to zero, and expansion parameter that is quite close to the Sedov phase. The age estimate is consistent within the uncertainties to the F06 values and the relatively low values of the reverse shock velocity and expansion parameter are the result of the relatively high ambient densities required by the $s=0$ case to match the data.

Note that velocities estimated from the optical proper motions are expected to lie between the ejecta velocity upstream of the reverse shock, $v_\mathrm{r,u}$, (less the cloud shock velocity of $\sim100\,\kms$) and the fluid velocity, $v_\mathrm{r,d}$, downstream of the reverse shock. The $v_\mathrm{r,d}$ applies to a diffuse component undergoing a Rankine-Hugoniot jump at the reverse shock: we interpret this component as producing the X-ray emission. Optical emission requires much slower shocks ($\sim 100\kms$ for [\ion{O}{3}] emitting shocks). A natural explanation is that the [\ion{O}{3}] emission results from ``cloud'' shocks driven into much denser material.

The properties of ``shock/cloud'' interactions were considered by \citet{mckee1975}, and the issue has been extensively studied since. In essence, for a large overdensity $\chi =  \rho_\mathrm{cloud}/\rho_\mathrm{ambient}$, the shock drives a slower shock into the cloud, and wraps around the cloud. The cloud expands laterally (increasing drag); the shock is subject to Rayleigh-Taylor and Richtmyer-Meshkov instabilities, and the sides are subject to Kelvin-Helmholtz instabilities. Eventually, the cloud is shredded and merges with the general postshock flow. The high density material initially moves ballistically, then slows, is shredded, and merges with the post reverse-shock flow. This implies that the velocities based on optical proper motions should lie between the initial free expansion (ballistic) velocity ($v_\mathrm{r,u}$) and the eventual post-reverse-shock flow ($v_\mathrm{r,d}$). Such a scenario was presented by \cite{patnaude2014} to explain the correlation and lack of correlation between the optical and X-ray emission in Cas~A.  Our estimates of the $v_\mathrm{r,u}$ and $v_\mathrm{r,d}$ are included in Table~\ref{tab:model.error}. These values are consistent with the velocity estimates of F06 for both the $s=0$ and $s=2$ cases, although the $s=0$ case is closer to the F06 result.

The explosion energies for the $s=2$ case are low for an ejecta mass of $2\,\msol$ but increase to $10^{51}\,\mathrm{erg}$ for an ejecta mass of $6\,\msol$.  Likewise, the ambient medium densities grow from 
0.22 to 0.66 $\mathrm{amu\,cm}^{-3}$ as the ejecta mass increases. The latter value is more consistent with our estimate of the ambient medium density from the X-ray spectral fits of $1.21~\mathrm{amu\,cm}^{-3}$ with its at least factor of two uncertainties. The swept-up mass and the core mass also increase as the ejecta mass increases resulting in large implied masses for the progenitor.  For example, an ejecta mass of $6\,\msol$ suggests a progenitor of $\sim60\,\msol$.  The unshocked mass estimates are all relatively low, less than $0.2\,\msol$.  One explanation for the fact that there is no or very little X-ray emitting Fe observed in the ejecta emission from E0102 is that the majority of the Fe has not been heated by the reverse shock yet.  \cite{nomoto1997}
and \cite{sukhbold2016} predict Fe ejecta masses of less than $0.12\,\msol$ for progenitors of $20\,\msol$
or more, so our estimate is consistent with this but implies that most of the unshocked ejecta is Fe. 
The unshocked ejecta also includes at least the O seen in the optically emitting [\ion{O}{2}] and [\ion{O}{3}] filaments.
Overall the model results for the $s=2$ case favor ejecta masses on the high end of our adopted range with the important implication that the progenitor mass would have had been close to the most massive stars in the SMC.

The explosion energies for the $s=0$ case are close to $10^{51}\,\mathrm{erg}$ for the lower ejecta masses
in Table~\ref{tab:model.results} but grow to $2.6\times10^{51}\,\mathrm{erg}$ for an ejecta mass of $6\,\msol$. The ambient medium densities for the low ejecta masses are more consistent with our estimate from the X-ray spectral fits.  The swept-up mass, the core mass, and the unshocked mass all show an increase with ejecta mass as in the $s=2$ case. Overall the model results for the $s=0$ case favor lower ejecta masses.  However, the estimates of the unshocked mass are less than $0.1\,\msol$, possibly inconsistent with the amount of Fe expected to be produced in such an explosion.  This inconsistency would only increase if the O plasma emitting optically in [\ion{O}{3}] and [\ion{O}{2}] were to be accounted for.

 Implicit in the modeling are assumptions about the evolution of the progenitor. Based on a detailed spectral analysis, \citet{blair2000} argued that the progenitor went through a Wolf-Rayet phase prior to core collapse. Our models are designed to fit both the measured and derived quantities, namely the remnant size, blastwave velocity, and ambient medium density at the blastwave. If the mass loss is assumed to be isotropic, then the ambient medium density of $\lesssim$ 1.0 cm$^{-3}$ suggests a high mass loss rate for the progenitor. If the progenitor exploded as a red supergiant, then the ambient density implies a mass loss rate $\lesssim$ 10$^{-4}$ M$_{\sun}$ yr$^{-1}$, for typical wind speeds of 10--20 km s$^{-1}$. If, on the otherhand, the progenitor exploded as a Wolf Rayet star as suggested by \citep{blair2000}, then mass loss rates of $\sim$ 10$^{-2}$ M$_{\sun}$ yr$^{-1}$ are required (for typical Wolf Rayet wind velocities of 1000 km s$^{-1}$), in order to match the blastwave velocity and radius. Unfortunately, with our current understanding of massive star evolution, neither of these possibilities seem plausible. \citet{smith2014} notes that for both red supergiants and Wolf-Rayet stars with solar metallicity, the mass loss rates do not extend much past 10$^{-5}$
M$_\sun$ yr$^{-1}$, and the mass loss from line driven winds scales as (Z/Z$_\sun$)$^{0.5}$, so maximum mass loss rates will be further reduced in the low metallicity environment of the SMC. 

When considering either isotropic mass loss scenario, neither is compatible with our current understanding
of massive star evolution. While detailed modeling of the progenitor's evolution would be required, it's plausible that the progenitor went through an extended Wolf Rayet phase prior to core collapse. For typical mass loss parameters of Wolf Rayet stars, a wind blown cavity and swept up shell of radius $\sim$ 6--7 pc will form on timescales of $\sim$ 20,000 years inside the red supergiant wind. In this scenario, the blastwave has only recently run into the cavity wall, and is now beginning to interact with the shell of swept-up red supergiant wind. The proposed model is supported by recent observations which show optically bright nebulosity exterior to the blastwave \citep{finkelstein2006,vogt2017}, which may be associated with the swept-up RSG wind, but would require detailed stellar evolutionary modeling, beyond the scope of this paper, to confirm.

The models we have used based on the descriptions in \cite{truelove99} and \citet{micel2016} with $s=0$  and $s=2$ profiles are simplifications of the mass loss history of the progenitor before explosion. The assumption of a steady stellar wind is at odds with recent work on episodic mass loss and binary mass transfer effects \citep[see][for a discussion]{smith2014}. The true mass loss most likely varied in time, may have had episodes of eruptive mass loss for relatively short periods of time, and most likely was highest towards the end of the life of the star.  If the progenitor was in a binary, the situation could be even more complicated. This variable mass loss is not captured in the model results we have presented.  These model results should be interpreted as a range of possibilities for the progenitor and initial conditions for the surrounding medium.  More detailed models and more restrictive constraints from the observations will be required to reduce the possible range suggested by these models.

\subsection{Electron-Ion Temperature Equilibration}
\label{subsec:e.i.temp.equil}
The counts weighted average electron temperature from our spectral fits is ${kT_{e}}\,\mathrm{(keV)}=0.68^{+0.05}_{-0.05}$. Our measured shock velocity can be used to estimate the temperature of the shocked gas: $kT = (3/16) \mu m_\mathit{p} v_\mathit{b}^2$ where $\mu \approx 0.61$ is the mean mass per free particle evaluated for cosmic abundances. With our measured shock velocity and its uncertainty, the estimated post-shock temperature of all particles is: $kT=3\mu m_{p}v_{b}^{2}/16=3.1_{-1.3}^{+1.6} \,\mathrm{keV}$. The difference between the electron temperature derived from the X-ray spectral fits and the post-shock mean temperature derived from the estimated shock velocity
indicates that the electrons and ions are not in equilibrium. Our spectral extraction regions of $1\arcsec$ in the radial direction correspond to a distance of $~8.8\times10^{17}\,\mathrm{cm}$. The time required for the shock to traverse a region of this length, assuming our estimate of the shock velocity, is $\Delta t = 5.5\times10^{9}\,\mathrm{s}$ (173 yr). Considering the gas after the shock is expanding with $3v_{b}/4$, the gas at the inner-most radii of our spectral extraction regions was shocked $4\Delta t = 2.2\times10^{10} \,\mathrm{s}$ ago. 
If we consider initial non-equilibration of the ion and electron  temperatures, the initial proton temperature immediately behind the shock is $T_\mathit{p,0}=5.09_{-2.05}^{+2.58}\,\mathrm{keV}$, and the electron temperature is $(m_\mathit{e}/m_\mathit{p})\,T_\mathit{p_0}$, assuming our estimate of the shock velocity.
Coulomb collisions will slowly equilibrate the temperature of the protons and electrons. From the pressure relation
we have \citep{itoh1978} 
   $n_\mathit{e} k T_\mathit{e} + n_\mathit{i} k T_\mathit{i}
   = (n_\mathit{e} + n_\mathit{i}) kT$,
which provides a relation  
$kT = (n_\mathit{e} T_\mathit{e} + n_\mathit{i} T_\mathit{i})/(n_\mathit{e} + n_\mathit{i})$
between the mean postshock temperature, the ion (predominantly proton) temperature $T_\mathit{i}$, and the electron temperature $T_\mathit{e}$.  The mean temperature $T$ can be obtained from the shock velocity $v_\mathit{b}$ 
and an estimate for the electron temperature, $T_\mathit{e}$ can be
estimated from the spectral fits providing $kT_\mathit{e}$.
It is not possible to measure the ion temperature directly
with observations at CCD energy resolution, but the 
ion (proton) temperature can be estimated indirectly from the
mean temperature and the electron temperature.

To estimate the equilibration, we assume a plasma consisting of
electrons and protons in the postshock plasma: $n_{e}=n_{p}$.
The time evolution of the temperatures of electrons and protons can be calculated as
\citep[Eqs. (B7), (B8)]{laming2003}:
\begin{equation}
\frac{d T_\mathit{e}}{d n_\mathit{e}t}
  = -\frac{d T_\mathit{p}}{d n_\mathit{e} t} = 0.13\frac{T_\mathit{p}-T_\mathit{e}}{2T_{e}^{3/2}}
 \label{eqn:te_tp}
\end{equation}
where densities are in units of $\mathrm{cm}^{-3}$, time is in s 
and $T_{p}$ and $T_{e}$ are in units of K.  This results in a pair
of coupled differential equations, and we use Runge-Kutta integration
to solve these equations from $n_\mathit{e}t=0$ to the weighted ionization time scale obtained by spectral fitting in 
Table~\ref{tab:blastwavespec}. Coulomb collisions between protons and electrons could increase the mean electron temperature to $1.02_{-0.24}^{+0.23}\,\mathrm{keV}$, averaging from $n_\mathit{e}\,t = 0$ to the emission-weighted ionization timescale $n_\mathit{e}t=1.73\times10^{11}\,\mathrm{cm}^{-3}\,\mathrm{s}$. 
The post-shock gas cools by adiabatic expansion as it expands after being shocked ($TV^{\gamma-1}=const.$, $\gamma=5/3$ for monatomic gas) reducing the mean electron temperature. Taking the adiabatic expansion of the post shock gas of our spectral extraction regions into account, the mean electron temperature could be $0.84_{-0.20}^{+0.18}\mathrm{keV}$ after averaging from $n_\mathit{e}\,t = 0$ to the emission-weighted ionization timescale $n_\mathit{e}t=1.73\times10^{11}\,\mathrm{cm}^{-3}\,\mathrm{s}$.  This estimate of the electron temperature agrees to within $1\sigma$ of our estimate of the electron temperature from the X-ray spectral fits. Therefore, we find that equilibration through Coulomb collisions and cooling through adiabatic expansion are sufficient to reconcile the electron temperature derived from the X-ray spectral fits and the electron temperature estimated from our measurement of the shock velocity.

\section{conclusions}
\label{sec:conclusion}

We present the first direct measurement of the expansion of the forward shock of E0102 of $0.025\%\pm0.006~yr^{-1}$, based on multiple X-ray observations with \chandra over a 17 year period.  
Our expansion rate implies a forward shock velocity of $1.61\pm0.37\times10^{3}~\mathrm{km~s}^{-1}$.

We exploited the superb angular resolution of \chandra to extract X-ray spectra from 5 narrow, annular regions near the forward shock that are dominated by emission from swept-up material and are mostly free of ejecta emission. We fit these spectra with a {\tt vpshock} model. 
%
%
The counts-weighted electron temperature from these fits is 
${kT_{e}}\,\mathrm(keV)=0.68^{+0.05}_{-0.05}$
Based on the electron temperature derived from the shock velocity 
(and accounting for Coulomb equilibration) we estimate 
$1.02_{-0.24}^{+0.23}\,\mathrm{keV}$. Accounting in addition for
adiabatic expansion reduces the estimate to 
${kT_{e}}\,\mathrm{keV}=0.84_{-0.20}^{+0.18}$. The electron 
temperature based on the X-ray spectral fitting and the estimate
based on shock velocity (as modified by Coulomb equilibration and
adiabatic expansion) are consistent to within $1\sigma$ uncertainty.
We take our measured values of the forward shock radius ($R_b=6.34$~pc), the reverse shock radius ($R_r=4.17$~pc), and the forward shock velocity and compare them to simple, one-dimensional shock models for $s=0$ (constant density ambient medium) and $s=2$  (constant wind) profiles to constrain the ejecta mass, explosion energy, ambient density, swept-up mass, and unshocked mass.  We find that models for both assumed profiles can reproduce the observed forward shock radius, reverse shock radius, and shock velocity but with significantly different values for the derived parameters.  Assuming that the explosion energy was between $0.5$ and $1.5\times10^{51}\,\mathrm{erg}$, the $s=2$ models prefer ejecta masses of $3-6\,\msol$, ambient densities of $\rho_{0}=0.33-0.66\, \mathrm{amu\,cm}^{-3}$, and swept-up masses of $26-52\,\msol$. These values imply the progenitor mass was in the range of about $30-60\,\msol$.  The $s=0$ models prefer ejecta masses of about  $2-3.5\,\msol$, ambient densities of about $\rho_{0}=0.85-1.5\,\mathrm{amu\,cm}^{-3}$, and swept-up masses of about $22-40\,\msol$.  Both the $s=0$ and $s=2$ cases predict upstream velocities of the ejecta that are consistent with the measured velocities of the optical filaments.
We note the limitations of these simple models (in particular the assumption of a steady stellar wind) and suggest that more detailed models with more restrictive constraints from existing and future observations will be needed to determine the progenitor and ambient properties more precisely.

\acknowledgments
L.X. acknowledges support by the National Key R\&D Program of China (2016YFA0400800), the National Natural Science Foundation of China under grant 11503027, and the UCAS Joint PhD Training Program(UCAS[2015]37). T.J.G., P.P.P., and D.J.P. acknowledge support under NASA contract NAS8-03060 with the \chandra X-ray center. J.P.H. acknowledges support for X-ray studies of supernova remnants from NASA grant NNX15AK71G. The authors thank John Raymond for useful comments on the manuscript.

\facilities{CXO}
\software{CIAO, FTOOLS, Xspec, SAOImage:DS9, Matlab}

\clearpage

\bibliographystyle{yahapj}
\bibliography{references}

\end{document}